\documentclass[twocolumn,showpacs,amsmath]{revtex4}
\usepackage{graphicx,amsmath,amsfonts, amssymb}
\newcommand{\Journal}[4]{#1 \textbf{#2}, #3 (#4)}

\newcommand{\PTO}{PbTiO$_3$}
\newcommand{\PZO}{PbZrO$_3$}
\newcommand{\BTO}{BaTiO$_3$}

\begin{document}

\title{ Hybrid exchange-correlation functional for accurate prediction of the electronic and structural properties of ferroelectric oxides}

\author{D. I. Bilc$^1$, R. Orlando$^2$, R. Shaltaf$^3$, G.-M. Rignanese$^3$, Jorge \'I\~niguez$^4$ and Ph. Ghosez$^1$}
\affiliation{
$^1$ Physique Th\'eorique des Mat\'eriaux, Universit\'e de Li\`ege (B5), B-4000 Li\`ege, Belgium 
\\
$^2$ Dipartimento di Scienze e Tecnologie Avanzate, Universit\`a del Piemonte Orientale, I-15100 Alessandria, Italy
\\
$^3$ Unit\'e de Physico-Chimie et de Physique des Mat\'eriaux, Universit\'e Catholique de Louvain, B-1348 Louvain-la-Neuve, Belgium
\\
$^4$ Institut de Ciencia de Materials de Barcelona (ICMAB-CSIC), 08193 Bellaterra, Spain}


\begin{abstract}

Using a linear combination of atomic orbitals approach, we report a systematic comparison of various Density Functional Theory (DFT) and hybrid exchange-correlation functionals for the prediction of the electronic and structural properties of prototypical ferroelectric oxides. It is found that none of the available functionals is able to provide, at the same time, accurate electronic and structural properties of the cubic and tetragonal phases of \BTO\ and \PTO. Some, although not all, usual DFT functionals predict the structure with acceptable accuracy, but always underestimate the electronic band gaps. Conversely, common hybrid  functionals yield an improved description of the band gaps, but overestimate the volume and atomic distortions associated to ferroelectricity, giving rise to an unacceptably large $c/a$ ratio for the tetragonal phases of both compounds. This {\sl super-tetragonality} is found to be induced mainly by the exchange energy corresponding to the Generalized Gradient Approximation (GGA) and, to a lesser extent, by the exact exchange term of the hybrid functional. We thus propose an alternative functional that mixes exact exchange with the recently proposed GGA of Wu and Cohen [Phys. Rev. B {\bf 73}, 235116 (2006)] which, for solids, improves over the treatment of exchange of the most usual GGA's. The  new functional renders an accurate description of both the structural and electronic properties of typical ferroelectric oxides.

\end{abstract}

\maketitle

\section{Introduction}

Ferroelectric (FE) oxides constitute an important class of multifunctional compounds, attractive for various technical applications in fields such as electronics, electromechanical energy conversion, non-linear optics or non-volatile data storage.\cite{Waser, Scott} Since the early 1990's, these compounds were the subject of numerous first-principles studies based on Density Functional Theory (DFT), most of which were performed within the usual Local Density Approximation (LDA) and, to a lesser extent, the Generalized Gradient Approximation (GGA).\cite{Rabe-Ghosez,Ghosez06} Although many successes have been achieved, the typical inaccuracies of these popular approximations impose considerable limitations. 

The LDA is well known to underestimate the lattice constants. While the typical underestimation of only 1 or 2 \% of the experimental value is in many contexts regarded as acceptable, it reveals itself quite problematic in the study of FE oxides whose structural instabilities are extremely sensitive to volume changes and, usually, suppressed significantly for the LDA equilibrium lattice constants.\cite{Rabe-Ghosez} In some cases, calculations are performed at the experimental volume to obviate this problem. However, such a method is not fully consistent and is unfeasible when experimental data are unavailable. Also, in the study of epitaxial multilayers in which different materials alternate, because the LDA errors pertaining to the lattice constants of the different compounds may be slightly different, it is impossible to impose a correct epitaxial strain simultaneously to all the layers.\cite{Zimmer02, Junquera03}

In most cases, the GGA constitutes a significant improvement over the LDA, although it has a tendency to overcorrect the LDA error for the equilibrium volume, thus leading to overestimations. In the case of ferroelectrics, however, the well known GGA functional of Perdew, Burke and Ernzerhof (GGA-PBE)~\cite{PBE} performs significantly worse than the LDA, yielding a wrong super-tetragonal structure in \BTO\ and \PTO.\cite{Wu2004} It was only very recently that Wu and Cohen~\cite{Wu2006} proposed a modified GGA functional (GGA-WC)  that is accurate for \BTO\ and \PTO\, and thus opens the door to a straightforward and reliable description of the structural properties of FE oxides. The so-called Weighted Density Approximation (WDA)~\cite{Gunnarsson79} has also been shown to constitute an improvement over the LDA.\cite{Singh97,Wu2004} However, since this approach is not available in the most widely used DFT codes, it has been applied only marginally.

It is also well known that the LDA and GGA usually lead to a significant underestimation of the band gaps of the Kohn-Sham electronic band structure (often by as much as a 50~\%).\cite{Martin04}  This does not {\it a priori} constitute an intrinsic failure: Kohn-Sham particles are fictitious independent particles resulting from a mathematical construction, with no other formal link to the real interacting electrons of the system than the requirement to reproduce the same total density, and no guarantee to exhibit the same energy spectrum. However, although ``exact'' Kohn-Sham DFT should internally correct for the so-called band-gap problem in order to provide accurate ground-state properties, that is not necessarily the case for approximate functionals. For example, the optical dielectric constant of insulators is usually badly reproduced within the LDA and GGA due to the absence of ultra non-local dependence of the exchange-correlation kernel.\cite{Gonze1995,Ghosez1997} The band-gap problem becomes even more critical in, for example, the emerging field of magnetic ferroelectrics, where the LDA and GGA often render erroneous predictions of metallicity for systems that are actually FE insulators~\cite{Hill1998,Hill2002,Shishidou2004}, thus precluding further investigations. Also, pathological situations may occur in the study of metal/FE interfaces: Because of the band-gap underestimation, the LDA often predicts the Fermi level of the metal to be erroneously aligned with the conduction bands of the FE, instead of being located within the gap. As a consequence, the LDA renders an unrealistic charge transfer from the metal to the FE, which in turn results in incorrect predictions for the properties of the interface or FE thin film under study. In view of all these failures, there is clearly a need for improved alternative functionals.

The so-called ``hybrid'' functionals~\cite{Becke1} that combine Hartree-Fock (HF) and DFT, such as B3PW~\cite{Becke1993}, B3LYP~\cite{CRYSTAL2005} and B1,~\cite{Becke1996,Ernzerhof1996} are very popular in Quantum Chemistry as they have been shown to provide accurate description of the atomization energies, bond lengths, and vibrational frequencies, together with good energy spectra for most molecules.\cite{Barone1994,Bausch1995,Baker1995,Tozer1996,Neumann1996,Dori2006,Riley2007} In the last decade, these hybrid functionals have been also increasingly applied to solids.\cite{Martin1997, Bredow2000, Muscat2001, Feng2004, Cora2004, Paier2006, Tran2006, Paier2007, Gerber2007} As it was for instance illustrated on Si, B3LYP constitutes an interesting alternative to LDA and GGA, as it provides excitation energies in much better agreement with the experiment.\cite{Muscat2001} Further, B3LYP has also been shown to significantly improve over the LDA results for magnetic oxides.\cite{Ruiz2003, Feng2004, Franchini2005,  Crespo2006, Franchini2007} Such hybrid functionals have been recently applied to ferroelectrics and, indeed, they provide a much better description of the electronic band gaps.\cite{Piskunov2004} However, a more exhaustive analysis that includes investigation of the structural and FE properties is still missing. 

In this paper, we consider two prototypical ferroelectrics, barium titanate (BaTiO$_3$) and lead titanate (PbTiO$_3$), and show that, although B3LYP and B1 give an acceptable description of the electronic properties, they fail to reproduce the correct tetragonal polar phase of these two compounds due to the super-tetragonality introduced by the GGA exchange part and, to a lesser extent, the exact exchange contribution. Then, we resort to the GGA recently introduced by Wu and Cohen, which is known to be accurate for the structural properties of ferroelectrics, and construct an alternative hybrid functional (B1-WC) that, while remaining accurate for the structural properties, is also accurate  in the prediction of the band gap and electronic structure. This functional opens the door to an improved theoretical description of both the electronic and structural properties of this important class of materials, as required in the especially challenging cases mentioned above. 

The outline of the paper is as follows. In Section II we briefly introduce the concept of hybrid functional. In Section III we describe the technical details of our calculations.  In Section IV we report the results obtained using the available functionals (LDA, GGA, B3LYP, B1 and HF), for the different phases of \BTO\ and \PTO\, and show that none of them gives simultaneously good structural and electronic properties. In the second part of the paper, we construct a different B1-WC hybrid functional by mixing the exact exchange with the GGA-WC functional. This functional is shown to provide an accurate description of both the electronic and structural properties of the ferroelectrics considered.

\section{Hybrid functionals}

DFT is {\sl a priori} an exact theory but relies, within the Kohn-Sham formalism, on an universal exchange-correlation energy functional, $E_{XC}$, whose explicit form is unknown and that must be approximated. In the search of accurate exchange-correlation Kohn-Sham functionals, an additional difficulty arises from the fact that $E_{XC}$, obtained from the integration of the exchange-correlation hole, is not equal to the exchange-correlation energy of the many-electron interacting system. The difference originates in the transfer of part of the many-body kinetic energy $T_{XC}$ to the exchange-correlation term within the Kohn-Sham formalism. This additional kinetic contribution makes $E_{XC}$ even more complex to estimate that it could be a priori expected.

An exact relationship is however satisfied by $E_{XC}$, known as the {\sl adiabatic connection formula} (or {\sl coupling-strength integration formula}).\cite{Harris1974} Let us consider the following family of Hamiltonians with different electron-electron interactions governed by a single parameter $\lambda$ varying from 0 (noninteracting limit) to 1 (fully interacting limit):
\begin{eqnarray}
H_{el}(\lambda) = \hat{T}_e + \lambda \hat{U}_{ee} + {v}_{\lambda}
\end{eqnarray}
where $\hat{T}_e$ is the kinetic energy operator, $\hat{U}_{ee}$ is the electron-electron potential energy operator and $v_{\lambda}$ is defined in such a way that all these Hamiltonians produce the same ground-state density {\it n}. For $\lambda = 1$, we have the fully interacting system with $v_{\lambda=1} = v_{ext}$, the external potential, while for $\lambda = 0$, we have the non-interacting system with $v_{\lambda=0} = v_{KS}$, the Kohn-Sham potential associated to the density {\it n}. It can be shown that the Kohn-Sham exchange-correlation energy corresponds to the average of the exchange-correlation potential energy for $\lambda$ ranging from 0 to 1,  the integration over $\lambda$ generating the kinetic part of $E_{XC}$, which is known as the {\sl adiabatic connection formula} ~\cite{Harris1974, Becke1988a}:
\begin{eqnarray}
\label{ACF}
E_{XC}=\int_0^1 d\lambda E_{XC,\lambda}.
\end{eqnarray} 
$E_{XC,\lambda}$ is the potential energy of exchange-correlation at intermediate coupling strength $\lambda$,
\begin{eqnarray}
\label{EXClambda}
E_{XC,\lambda}=\langle \psi_{\lambda} \mid \hat{U}_{ee} \mid \psi_{\lambda} \rangle - E_H
\end{eqnarray}
where $E_H$ is the classical electrostatic Hartree energy and $\psi_{\lambda}$ are the ground-state wave functions at the coupling strength $\lambda$. 

\begin{figure}[t]
  \centering\includegraphics[angle=0,scale=0.5]{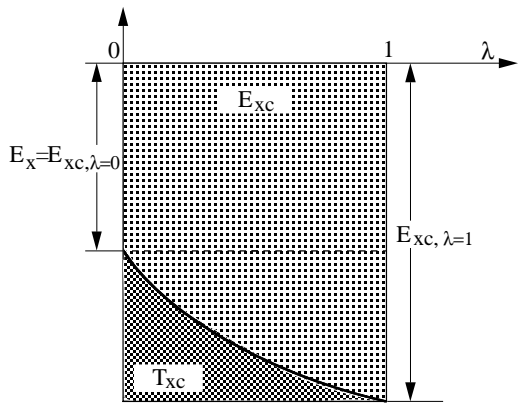}\\[-10pt]
   \caption{\label{Excfig} Kohn-Sham exchange-correlation energy $E_{XC}$ (light gray area ) is obtained as the potential energy of exchange-correlation of the truly interacting system $E_{XC,\lambda=1}$ modified by a positive quantity of the transfered kinetic energy $T_{XC}$ (dark gray area) all along the path of integration of the coupling constant $\lambda$.}
\end{figure}

This result is illustrated in Fig.~\ref{Excfig}. At $\lambda = 0$,  from Eq. \ref{EXClambda}, $E_{XC,\lambda=0} = E_X$  where  $E_X = \langle \psi_{0} \mid \hat{U}_{ee}
\mid \psi_{0} \rangle - E_H$ is the exact exchange energy of the system defined as
within the Hartree-Fock method but on the basis of the Kohn-Sham wavefunctions
$\psi_{0}$. At $\lambda =1$, $E_{XC,\lambda=1}$ in Eq. \ref{EXClambda}  corresponds to the many-body potential energy of exchange-correlation of the fully interacting system. The Kohn-Sham exchange-correlation energy, $E_{XC}$, is provided by Eq. \ref{ACF} and corresponds to the light gray area in Fig.~\ref{Excfig}. It differs from the many-body exchange-correlation energy of the fully interacting system by $T_{xc}$ (dark gray area in Fig.~\ref{Excfig}) corresponding to the transfer of many-body kinetic energy along the path of integration over $\lambda$ between 0 and 1.

 From Fig.~\ref{Excfig} it appears that $E_{XC}$ could be obtained from mixing of exact exchange energy ($E_X = E_{XC,\lambda=0}$) and many-body ($E_{XC,\lambda=1}$) exchange-correlation potential energy (end-points of the integration path over $\lambda$). The exact value of this mixing is however unknown and depends on the shape of the curve describing the evolution of $E_{XC,\lambda}$ with $\lambda$.  If the evolution was perfectly {\it linear}, we would simply get the exact result,
\begin{eqnarray}
\label{mean}
E_{XC} = \frac{1}{2} (E_X+E_{XC,\lambda=1})
\end{eqnarray}

The simple half-and-half hybrid functional proposed by Becke is based on Eq.~\ref{mean} and it is defined as~\cite{Becke1,Levy1996}: 
\begin{eqnarray}
E_{XC}^{hyb} = \frac{1}{2} (E_X+E_{XC,\lambda=1}^{LDA}).
\end{eqnarray}
More sophisticated mixing schemes were further proposed to better capture the evolution of $E_{XC,\lambda}$ with $\lambda$.

B3PW and B3LYP are based on the following mixing scheme :
\begin{equation}
\label{Eq1}
\begin{split}
E_{XC}^{B3LYP} = E_{X}^{LDA} + A(E_{X} - E_{X}^{LDA}) + (1-A)B(E_{X}^{GGA} \hspace{0.1 in} \\ -E_{X}^{LDA})
+ E_{C}^{LDA} + C(E_{C}^{GGA} - E_{C}^{LDA}), \hspace{0.4 in} 
\end{split}   
\end{equation}   
where A=0.2, B=0.9, C=0.81 are the three Becke's mixing parameters which were determined by fitting  experimental data on molecules~\cite{Becke1993}, $E_{X}^{LDA}$ and $E_{C}^{LDA}$ are the exchange and correlation energy within LDA, and $E_{X}^{GGA}$ and $E_{C}^{GGA}$ are the exchange and correlation energy within the GGA functional. The B3PW functional includes the Becke's GGA exchange~\cite{Becke1988} and the GGA correlation of Perdew and Wang~\cite{Perdew1991}, whereas the B3LYP functional includes the Becke's GGA exchange~\cite{Becke1988} and the GGA correlation of Lee, Yang and Parr~\cite{Lee1988}  for $E_{X}^{GGA}$ and $E_{C}^{GGA}$ respectively. 

The more recent B1 hybrid functional sets B=1 and C=1 simplifying the mixing 
procedure to the only exact exchange mixing parameter A~\cite{Becke1996, Ernzerhof1996}:
\begin{equation}
\label{Eq2}
E_{XC}^{B1} =  E_{XC}^{GGA} + A  (E_{X} - E_{X}^{GGA})  
\end{equation}   
Fitting the experimental data, Becke determined the values of 0.16 and 0.28 for the A parameter depending on the choice of the GGA functional used in $E_{X}^{GGA}$ and $E_{C}^{GGA}$.\cite{Becke1996} Perdew, Ernzerhof and Burke provided a qualitative physical justification for the B1 functional and for the empirical value of A parameter.\cite{Perdew1996} They showed that A$\approx$0.25 is to be expected for the atomization energies of most molecules, while a larger values may be more appropriate for total energies of atoms and molecules, and smaller values for atomization energies of molecules with nearly degenerate ground states of the unperturbed $\lambda =0$ problem.

\section{Technical details}

Our calculations have been performed using the linear combination of atomic orbitals (LCAO) method as implemented in the CRYSTAL code.\cite{CRYSTAL2005} Results have been obtained using (i) DFT at the LDA and GGA levels, (ii) hybrid functionals B3LYP and B1, and (iii)  the Hartree-Fock method. For the LDA calculations, we used  the Dirac-Slater exchange~\cite{Dirac1930} and the Vosko-Wilk-Nusair correlation~\cite{Vosko1980} functionals. For the GGA calculations, we considered the exchange and correlation functional of Perdew, Burke and Ernzerhof (GGA-PBE)~\cite{PBE} as well as that of Wu and Cohen (GGA-WC).\cite{Wu2006} As for the hybrids, we considered the conventional B3LYP functional and a B1 functional with $A = 0.16$, using Becke's GGA exchange and Perdew-Wang GGA correlation functionals.\cite{Becke1996} A different B1-WC functional that we introduce in Section V-B makes use of the Wu-Cohen GGA exchange-correlation functional and $A = 0.16$. 

We should note here that our calculations for the various exchange-correlation functionals differ {\sl only} in the exchange-correlation functional used; that is, all the remaining parameters and approximations, including the pseudopotentials, are common to all our calculations. It should be noted here that keeping always the same pseudopotentials, irrespectively of the exchange-correlation functional, seems to be the dominant way to proceed in the community traditionally working with HF and hybrid functionals. The pseudopotentials employed are generated from HF calculations, and it can be argued~\cite{Piskunov2004} that they constitute a reasonable choice. On the other hand, the DFT community working on ferroelectrics has always favored an alternative approach that involves the generation of the corresponding pseudopotentials for each exchange-correlation functional. Being aware of these two possibilities, in this work we have performed all our calculations following two approaches: (i) we used HF pseudopotentials for Ba, Pb and Ti atoms, considering explicitly all the electrons of oxygen {\sl and} (ii) we have also performed calculations in which Ti, together with O, is treated at the all-electron level. Since, the Ti and O atoms play a key role as far as the FE properties are concerned, such an approach allows us to quantify the magnitude of the error that might be caused by the use of the common HF pseudopotentials. Interestingly, we will see that such a pseudopotential approximation is reasonable except in the case of the Wu-Cohen GGA functional. The HF pseudopotentials and the localized Gaussian-type basis sets used here are those from Ref.~\cite{Piskunov2004}. The basis sets include polarization $d$-orbitals for O ions and were optimized for both \BTO\ and \PTO. For the calculations of the cubic phases of other perovskite compounds which we have considered, the HF pseudopotentials and basis sets are as follow: Sr (from Ref.~\cite{Piskunov2004}), Ca (from Ref.~\cite{Catti1991}), Zr (all electron from Ref.~\cite{DovesiZrbasis}), K (all electron from Ref.~\cite{Dovesi1991}), and Ta (from Ref.~\cite{Bredow2006}).

Other technical details are as follows. Brillouin zone integrations were performed using a 6$\times$6$\times$6 mesh of $k$-points, and the self-consistent-field calculations were considered to be converged when the energy changes between interactions were smaller than 10$^{-8}$ Hartree. An extra-large predefined pruned grid consisting of 75 radial points and 974 angular points was used for the numerical integration of charge density. For the cubic phases we have optimized the lattice constants, whereas for the tetragonal phases we have performed, except where otherwise indicated, full optimizations of the lattice constants and atomic positions. The optimization convergence of 3$\times$10$^{-5}$ Hartree/Bohr in the root-mean square values of forces and 1.2$\times$10$^{-4}$ Bohr in the root-mean square values of atomic displacements was simultaneously achieved for both atomic position and lattice constant optimizations. The level of accuracy in evaluating the Coulomb and exchange series is controlled by five parameters.\cite{CRYSTAL2005} The values used in our calculations are 7, 7, 7, 7, and 14. For computing the spontaneous polarization a denser mesh of 10$\times$10$\times$10 $k$-points was used. Also, for phonon frequency calculations we have used the finite atomic displacements of 0.0005 \AA\ for numerical evaluations of the energy derivatives and the energy convergence was increased to 10$^{-12}$ Hartree.

The quasiparticle band structure in Section IV-A was calculated within the $GW$ approximation~\cite{Hedin} as implemented in the ABINIT code.\cite{ABINITGW} The LDA eigenvalues were obtained using separable, extended norm-conserving pseudopotentials~\cite{Teter} and used as zero-order input in a perturbative fashion.\cite{ Hybertsen-Godby} The LDA calculation was performed for a 4$\times$4$\times$4 $k$-point mesh inside the Brillouin zone with a cutoff energy of 50 Hartree. The dielectric matrix $\epsilon^{-1}_{\bf G,G'}({\bf q})$ was computed within the random phase approximation (RPA) for 461 $G$ vectors including 200 bands (20 occupied and 180 unoccupied) in the summations. The dynamic dependence of $\epsilon^{-1}$ was approximated using a plasmon-pole model~\cite{Godby-Needs} fitted to the actual calculated values of $\epsilon^{-1}$ at two imaginary frequencies.  The self-energy $\Sigma$ was obtained by summing over 19 special $k$-points in the irreducible Brillouin zone, and over 400 bands (20 occupied and 380 unoccupied). This procedure was repeated iteratively by correcting the unoccupied eigenvalues with a scissor operator, matching the quasiparticle gap from one iteration to construct $\epsilon^{-1}$ and the Green's function for the next iteration, until convergence of the quasiparticle gap was reached. The eigenstates were
assumed not to change. The accuracy of this procedure is beyond G0W0 in which no update is done at all and below self-consistent $GW$ in which both the eigenvalues and eigenstates are updated. Finally, let us note we checked that, in spite of the technical differences between the two calculation schemes (e.g. basis sets and pseudopotentials), ABINIT and CRYSTAL give essentially the same result for the electronic band structure of \BTO\ at the LDA level. It is thus meaningful to compare the density-functional results obtained with CRYSTAL with the $GW$ ABINIT results.

\begin{figure}[t]
  \centering\includegraphics[angle=0, scale=0.21]{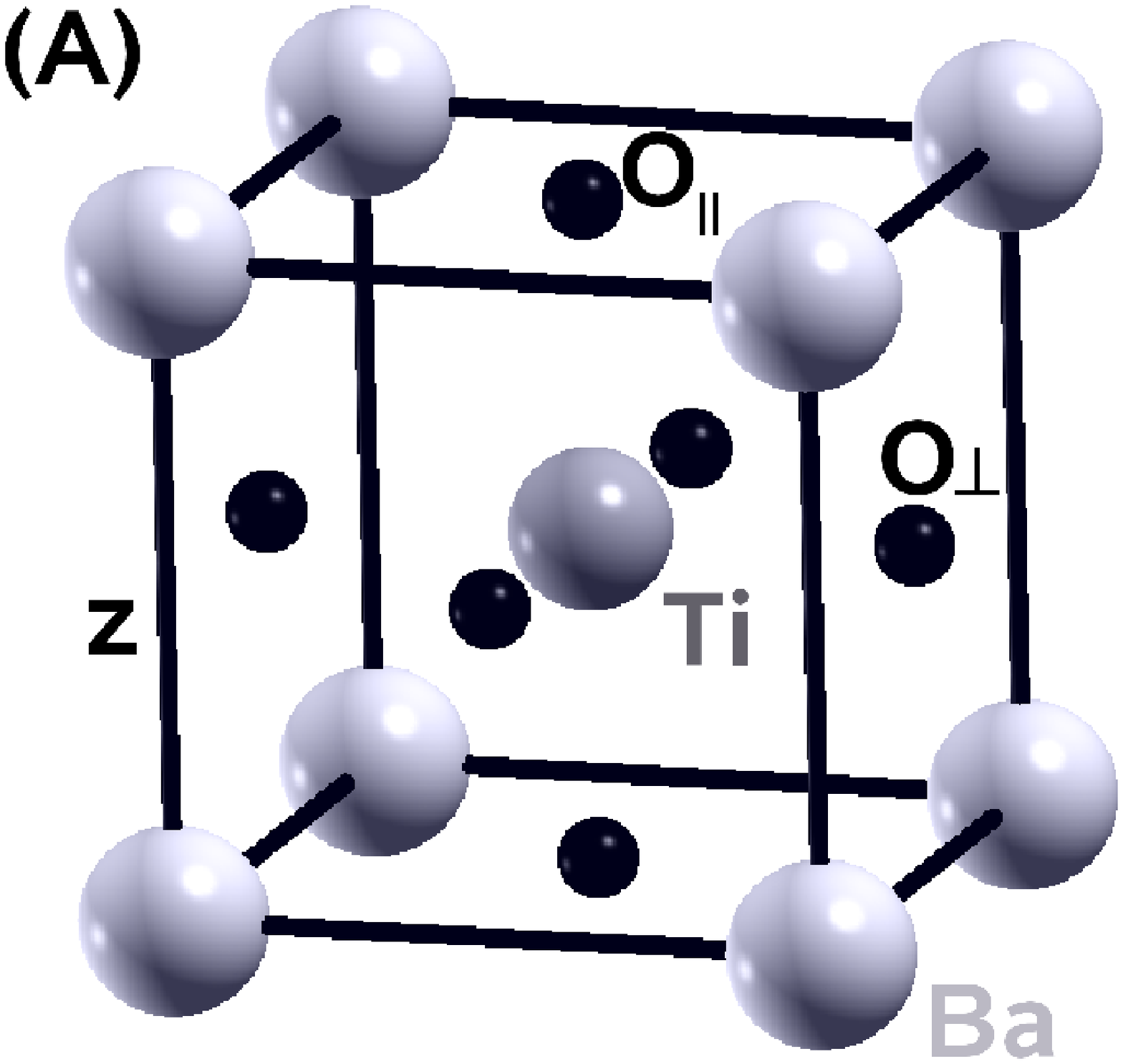}%
  \centering\includegraphics[angle=0,scale=0.21]{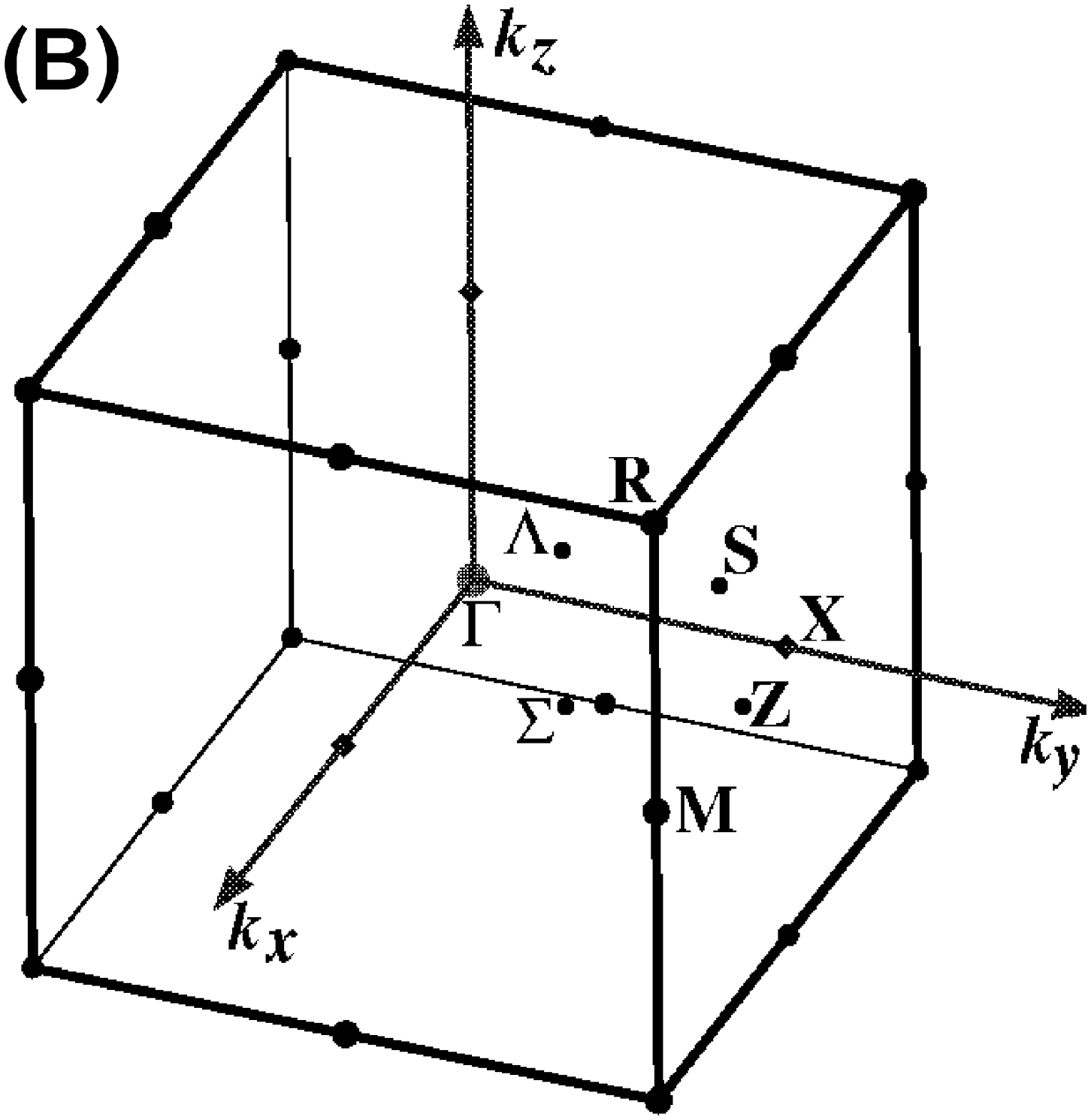}\\[-15pt]
   \caption{\label{Strucfig} Cubic perovskite unit cell (A) and the corresponding Brillouin zone (B) for BaTiO$_3$.}
\end{figure}

\section{Comparison of available functionals}

BaTiO$_3$ and PbTiO$_3$ are two prototypical FE perovskites. They both adopt at sufficiently high temperature a paraelectric $Pm{\overline 3}m$ cubic structure as shown in Fig.~\ref{Strucfig}. When the temperature is lowered, they both undergo a structural phase transition ($T_c = 393$ K and 766 K for BaTiO$_3$ and PbTiO$_3$, respectively) and exhibit a FE $P4mm$ tetragonal phase at room temperature. While PbTiO$_3$ remains tetragonal down to 0~K, BaTiO$_3$ undergoes two additional structural phase transitions when the temperature is lowered, to orthorhombic and rhombohedral phases, respectively. The present study focuses on the cubic and tetragonal phases of these two representative compounds.\cite{footnote} In this Section, we compare the ability of the available DFT (LDA, GGA-PBE, GGA-WC) and hybrid (B3LYP, B1) exchange-correlation functionals to predict correctly their electronic, structural and FE properties.\\[-20pt]

\subsection{Electronic band structure}

\begin{figure}[t]
  \centering\includegraphics[angle=-90, scale=0.26]{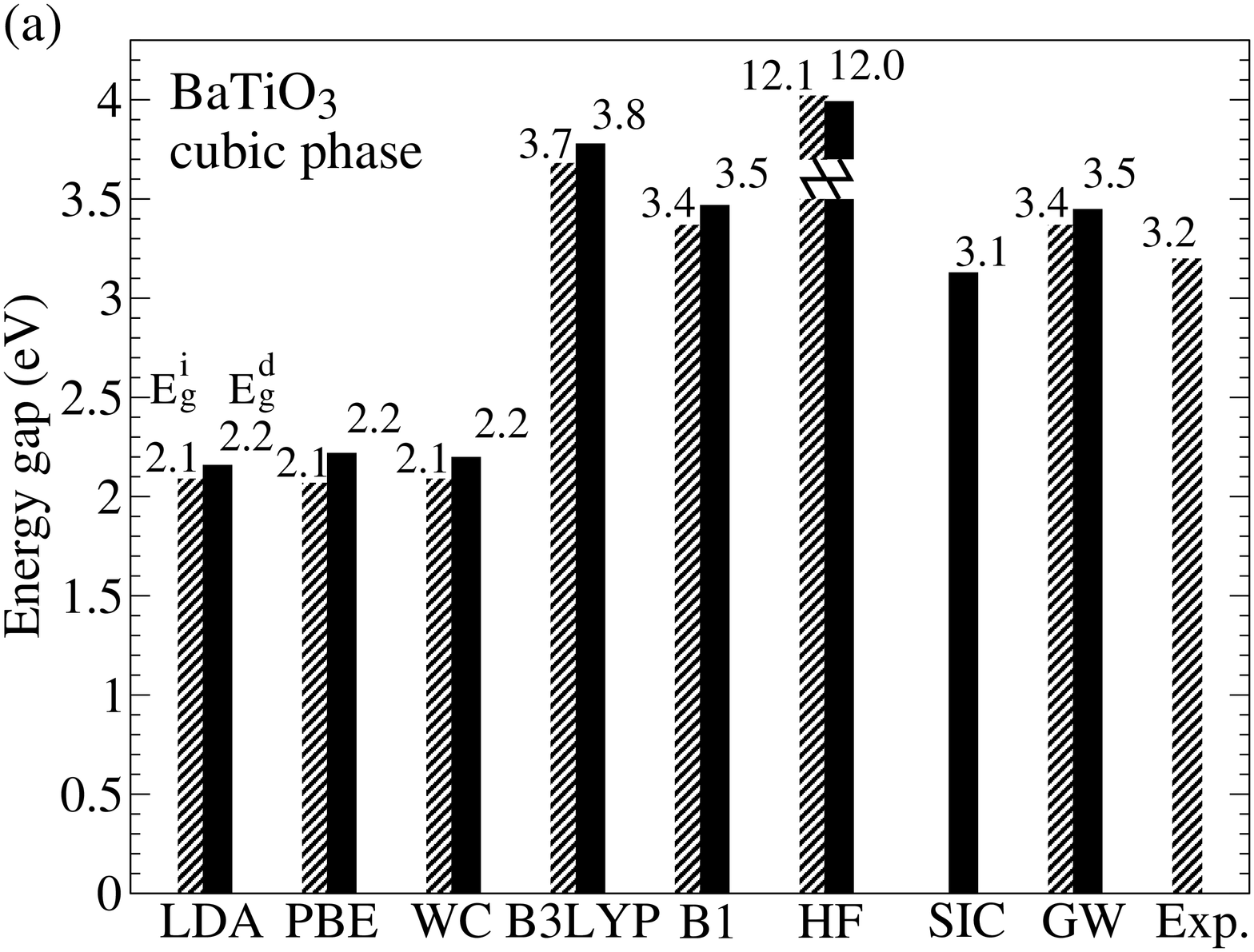}
  \centering\includegraphics[angle=-90,scale=0.26]{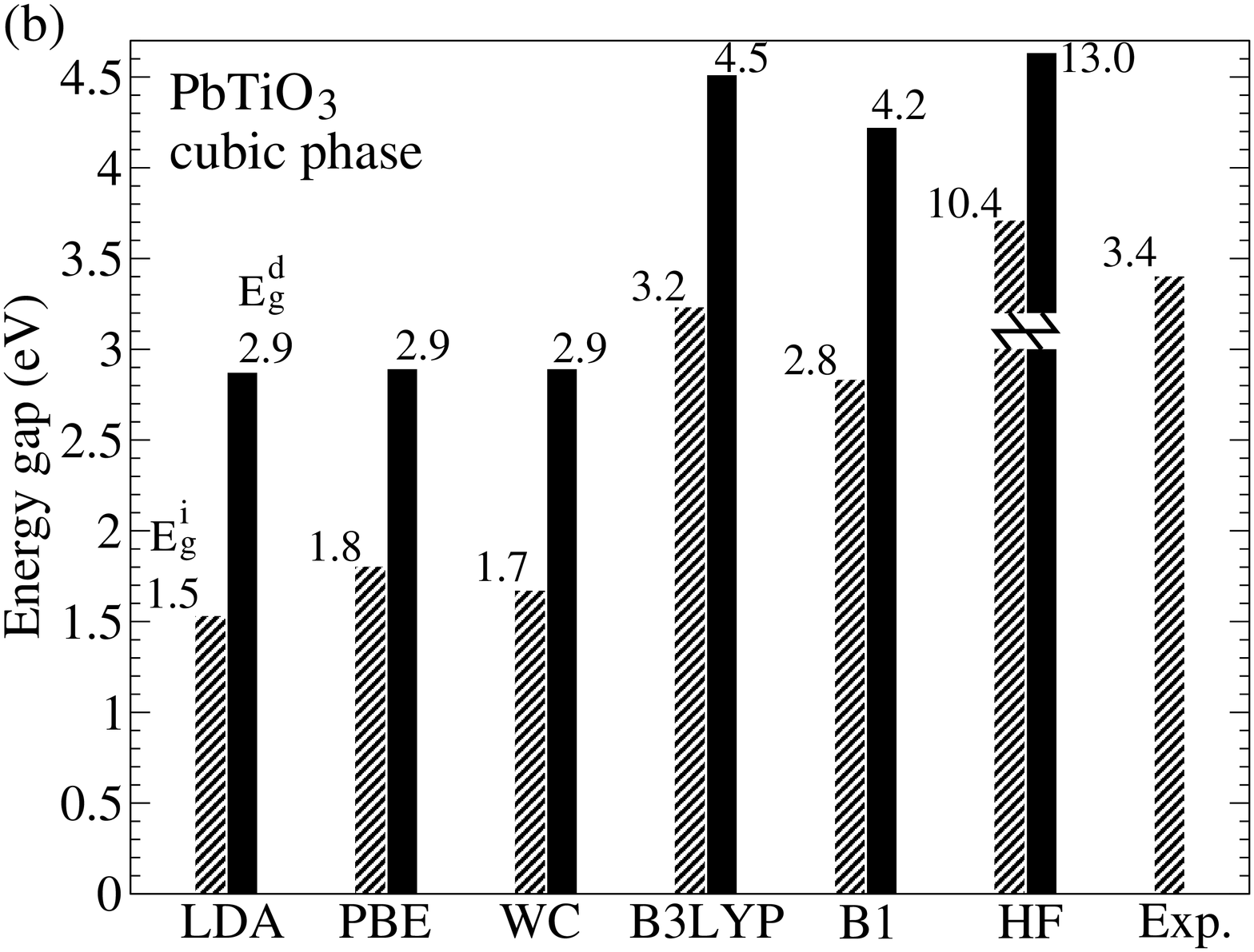}\\[-10pt]
   \caption{\label{Egfig} Indirect $E^i_g$ and direct $E^d_g$ band gaps for the Ti all-electron calculations within different functionals for: (a)\BTO\ and (b)\PTO. $E^i_g$ are shown in dashed oblique lines and $E^d_g$ in continuous lines. The experimental $E^i_g$ values are also shown.\cite{Wemple1970, Peng1992} The pseudo-SIC and $GW$ values for \BTO\ are also included.}
\end{figure}

First, we compare the results obtained with the available functionals for the electronic band structure. In Fig.~\ref{Egfig}, we report the indirect ($E^i_g$) and direct ($E^d_g$) band gaps for the cubic phases of \BTO\ and \PTO, as obtained for Ti all-electron calculations.\cite{note-1} The values were calculated at the corresponding theoretical lattice constants. For \BTO\ (resp. \PTO) $E^i_g$ occurs along [$R-\Gamma$] (resp. [$X-\Gamma$]) direction of the cubic Brillouin zone (Fig.~\ref{Strucfig}b). For \BTO, the results obtained with a recently proposed self-interaction--correction scheme (pseudo-SIC)~\cite{Pilippetti2003} and including $GW$ corrections are also shown. The experimental $E^i_g$ values for \BTO~\cite{Wemple1970} and \PTO~\cite{Peng1992} are finally mentioned for comparison. 

On the one hand, it is shown that both LDA and GGA (PBE and WC) strongly underestimate the band gaps giving rather similar values of $E^i_g$ and $E^d_g$ (Fig.~\ref{Egfig}). On the other hand, HF strongly overestimates these values. Only B3LYP and B1 hybrid functionals give improved $E^i_g$ and $E^d_g$ values as previously reported in Ref.~\cite{Piskunov2004}.\\[-20pt]
\begin{table}[h]
\begin{center}
\caption{\label{Bndshifttable} $GW$ corrections to the LDA eigenvalues for the bottom of conduction band ($\Delta E_{CB}$) and the top of valence band ($\Delta E_{VB}$)  at different high symmetry points in the cubic Brillouin zone.}
 \begin{tabular*}{0.48\textwidth}%
    {@{\extracolsep{\fill}}ccccc}
\hline\hline
                    & $\Gamma$ & $X$ & $M$  & $R$  \\
\hline
$\Delta E_{CB} (eV)$ & 1.01 & 1.05 & 1.03 & 1.08 \\
$\Delta E_{VB} (eV)$ & -0.54 & -0.59 & -0.55 & -0.54 \\
\hline\hline
\end{tabular*}
\end{center}
\end{table}    

\begin{widetext}
\begin{figure}[t]
\begin{minipage}[t]{1.0\textwidth}
  \centering\includegraphics[scale=0.305]{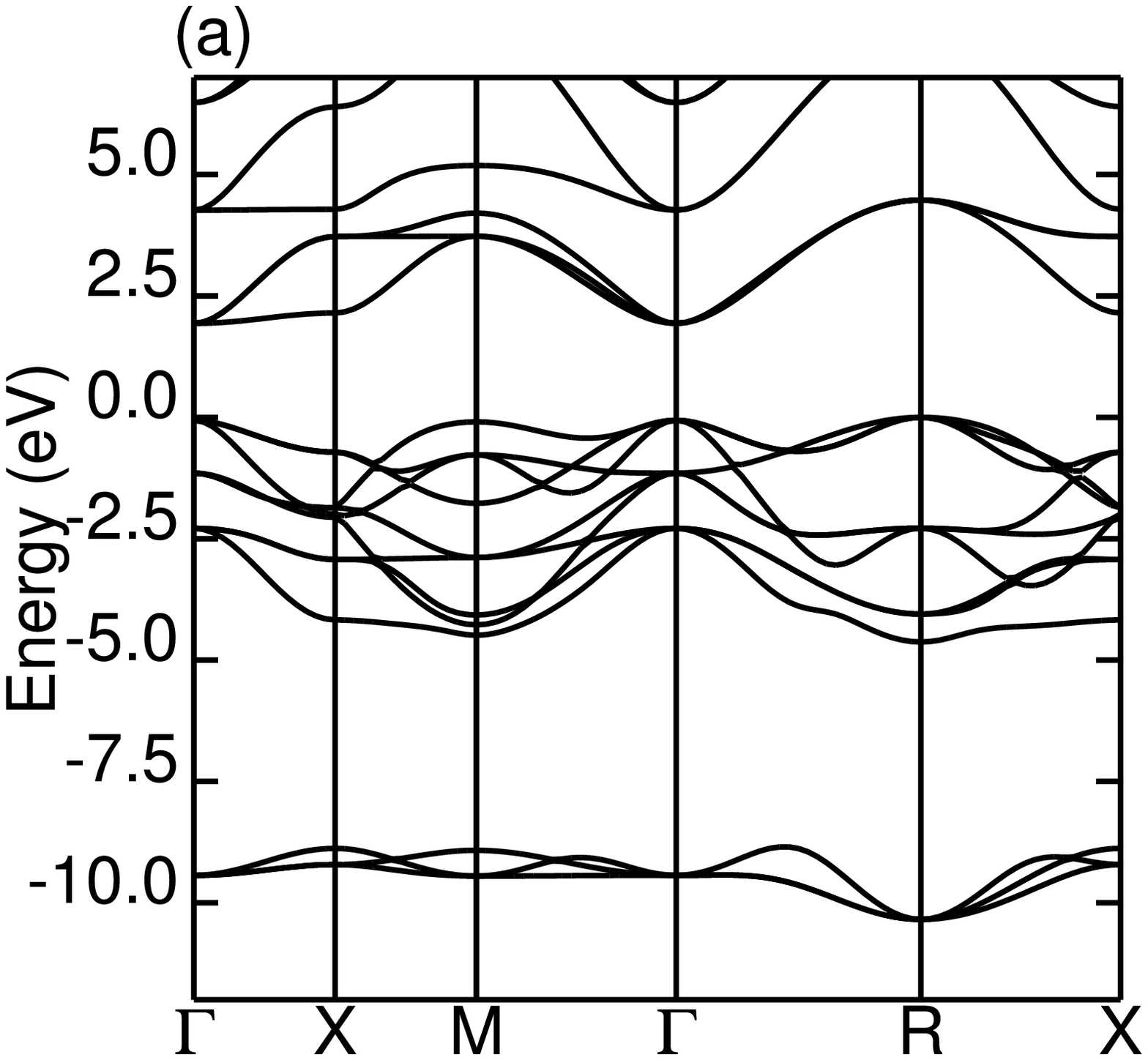}%
  \centering\includegraphics[scale=0.305]{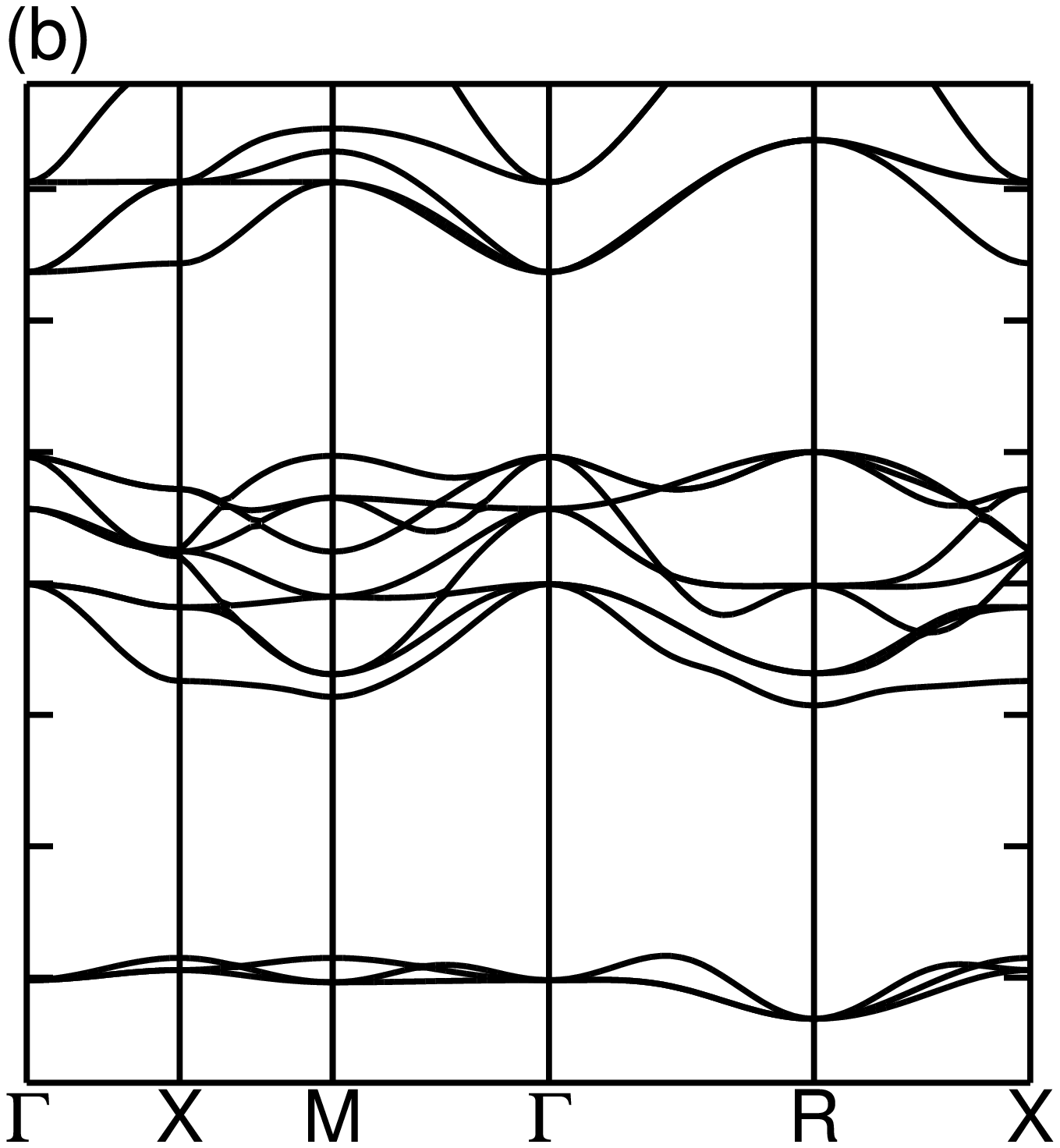}%
  \centering\includegraphics[scale=0.305]{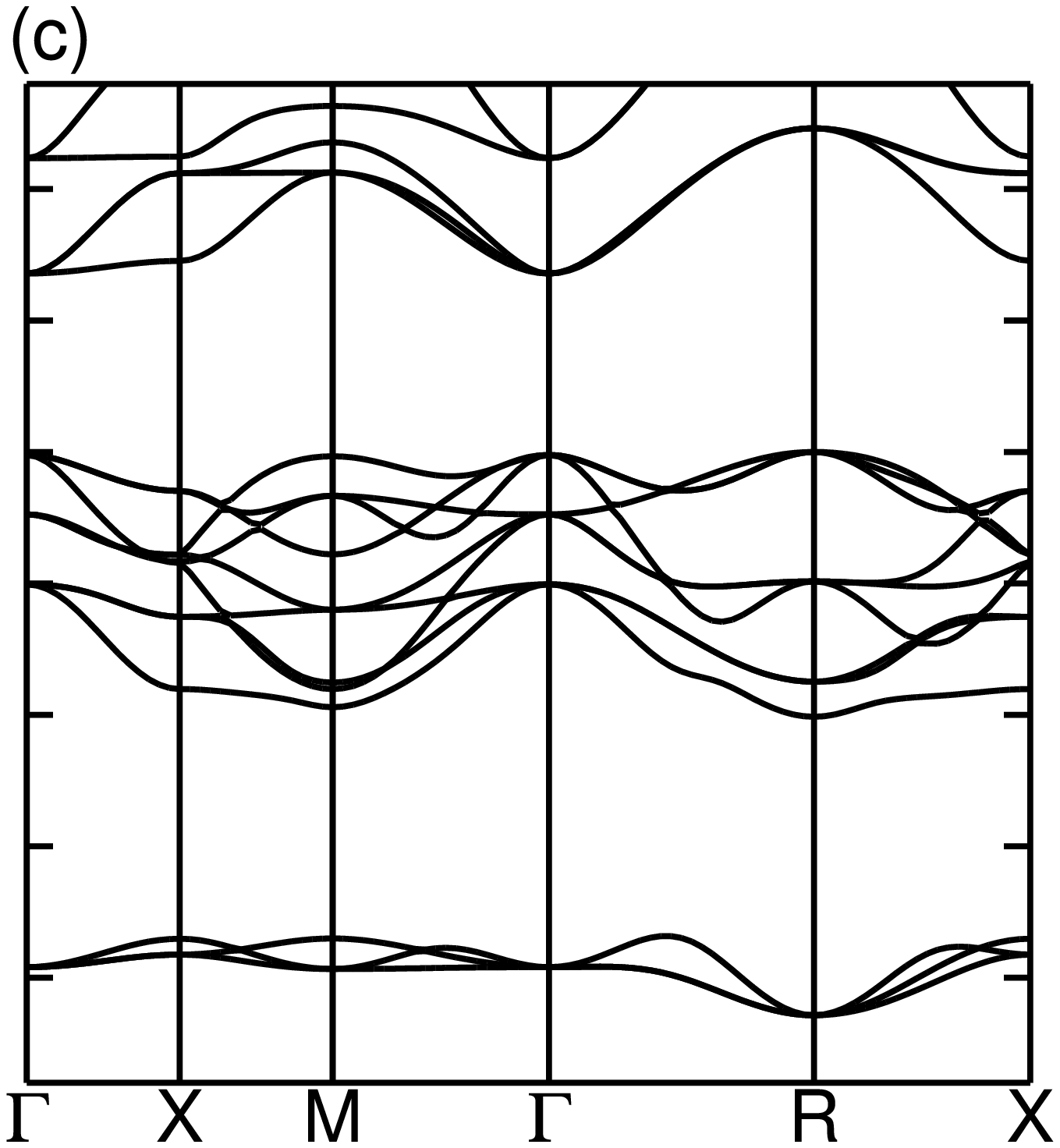}%
  \centering\includegraphics[scale=0.228]{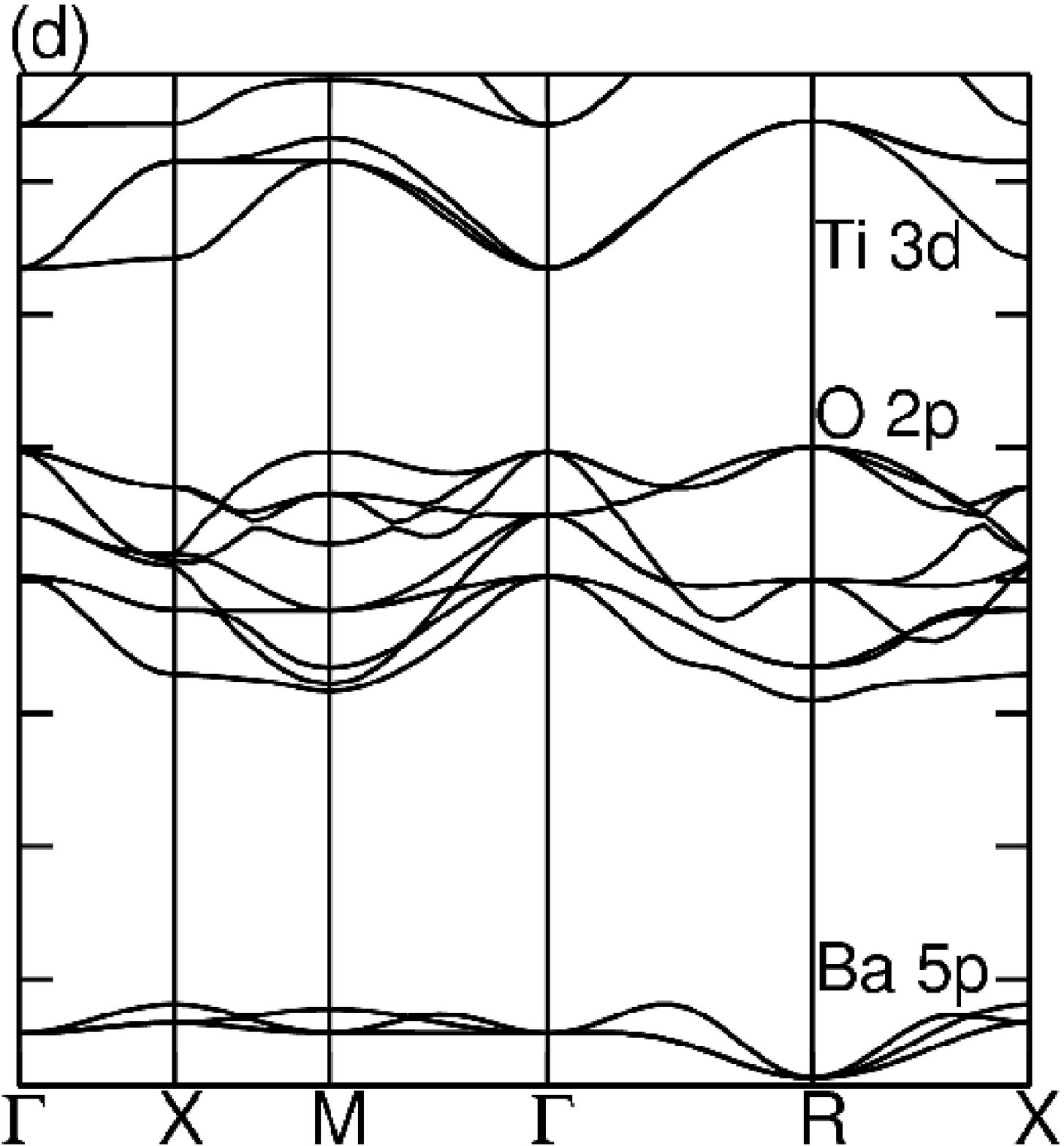}\\[-10pt]
  \caption{\label{Bndfig} Electronic band structure of cubic \BTO\ at theoretical lattice constants within: (a)LDA, (b)B3LYP, (c)B1-WC and (d)$GW$ approximation. The LDA, B3LYP and B1-WC band structures are given for Ti all-electron calculations.}
\end{minipage}
\end{figure}
\end{widetext}

It is interesting to discuss not only $E_g$ values but also the whole dispersion of the electronic band structure obtained within the different functionals. In Fig.~\ref{Bndfig}, we report the electronic band structure of cubic \BTO\ calculated within LDA, B3LYP and including $GW$ corrections. B1-WC results are also shown although they are only discussed in Section V-B. The 0~eV energy level corresponds to the top of the valence band (VB).

The group of valence bands in the energy range between $-$5 and 0~eV has a dominant O~2$p$ orbital character, whereas the VB levels localized around $-$11~eV are composed mainly of Ba~5$p$ orbitals (see the labels in Fig.~\ref{Bndfig}d). The first five conduction bands (CB's) correspond essentially to the Ti~3$d$ states and these split in two groups of  $t_{2g}$ and $e_g$ symmetry, respectively. The experimental results of the photoelectron spectroscopy of \BTO\ give a VB spread of $\sim5.5\pm0.2$~eV, Ba~5$p$ orbitals localized around $-$12$\pm0.2$~eV, and an $E_g$ value of $\sim3.3\pm0.2$~eV.\cite{Hudson1993} The $GW$ results are in good agreement with these data, taking into account the experimental uncertainties.   

The $GW$ correction to the LDA band structure does not simply correspond to a rigid upward shift of the Ti~3$d$ conduction bands. Table~\ref{Bndshifttable} shows the individual shifts of the bottom of CB ($\Delta E_{CB}=E_{CB}^{GW}-E_{CB}^{LDA}$) and the top of VB ($\Delta E_{VB}=E_{VB}^{GW}-E_{VB}^{LDA}$) at various high symmetry points of the cubic Brillouin zone. It highlights that the increase of the band gap is due both to an upward shift of the bottom of CB and a downward shift of the top of VB. The Ba~5$p$ VB states are also differently affected by the $GW$ correction and, as a result, their energy reduces with respect to that of the O~2$p$ states.

Fig.~\ref{Bndfig}b shows the B3LYP band structure, which is in much better agreement with the $GW$ band structure than the LDA result. In particular, B3LYP and $GW$ render a very similar dispersion of the O~2$p$ top VB states. A few noticeable differences between the B3LYP and $GW$ band structures exist such as (i) a near degeneracy at the X point in the CB of B3LYP which is lifted up in the case of $GW$, and (ii) the position of the Ba~5$p$ VB states which are higher in energy.

\subsection{Structural properties}

Next we investigated the structural properties using the different functionals. We performed full optimizations of the cubic and tetragonal phases. The results of these optimizations are shown in Table~\ref{Strtable}.

In the cubic phase, the lattice constants are in good agreement with those of Ref.~\cite{Piskunov2004}. Theoretical results from other authors are also shown for comparison. For \BTO, both the LDA and GGA-WC slightly underestimate the experimental lattice constant (4.00 \AA) whereas the GGA-PBE, B3LYP, B1 and HF slightly overestimate it. For \PTO, the value of the cubic lattice constant at the FE phase transition (766 K) is 3.969 \AA. If this value is extrapolated to 0 K it reduces to 3.93 \AA.\cite{Mabud1979} Taking this extrapolated value as the reference, then again the LDA and GGA-WC slightly underestimate the lattice constant whereas the GGA-PBE, B3LYP, B1 and HF slightly overestimate it. The values of the lattice constants obtained from Ti all-electron calculations are very close to those obtained using Ti pseudopotentials, the largest difference (0.006 \AA) corresponding to the HF case. In summary, these optimizations of the cubic phases show that LDA produces the largest deviation ($\sim 1\%$) compared to the experiment for both \BTO\ and \PTO\, while the GGA-WC, B1 and HF schemes give all very good values for the lattice constants.

As for tetragonal phase, all the functionals considered here give similar atomic distortions when the atomic positions are relaxed at fixed experimental lattice constants. (Wu {\it et. al.}~\cite{Wu2004} had already observed that this is the case for the LDA and GGA functionals.) The atomic distortions for all the functionals are comparable with the experiment for \BTO\ and slightly overestimated for \PTO.

The situation is different when cell parameters and atomic positions are relaxed together. The results are reported in Table~\ref{Strtable}. On the one hand, the LDA underestimates the volumes and atomic displacements as compared to experiment but gives good values for the tetragonality $c/a$. On the other hand the GGA-PBE, B3LYP, B1 and HF schemes overestimate the volumes and atomic displacements, leading to very large $c/a$ ratios for both \BTO\ and \PTO. Such a super-tetragonality had already been reported for the GGA-PBE functional in the case of \PTO.\cite{Wu2004} For \BTO (resp. \PTO), the largest overestimate in the tetragonality, $\sim 6$\% (resp. $\sim 19$\%), is given by the B3LYP (resp. B3LYP and HF). The values of the tetragonality given by all functionals within the Ti all-electron calculations are only slightly reduced as compared to those obtained from Ti pseudopotential calculations, with the exception of the GGA-WC results for \PTO. In that case, the tetragonality is strongly reduced from 1.131 to 1.086, a result that comes close to the experimental value of 1.071 extrapolated at 0~K.\cite{Mabud1979} Such a special sensitivity of the GGA-WC functional to the use of HF pseudopotentials deserves further analysis. In summary from Table~\ref{Strtable}, only the GGA-WC functional (and, to a lesser extent, the LDA) succeed in correctly predicting the right tetragonal structures of \BTO\ and \PTO.

\subsection{Ferroelectric properties}

Finally, we also investigated how the different available functionals reproduce typical FE properties such as the Born effective charges, the spontaneous polarization and the FE double-well energy.

\begin{widetext}
\begin{table}[t]
\begin{minipage}[t]{1.0\textwidth}
\begin{center}
\caption{\label{Strtable} Fully relaxed cubic and tetragonal lattice constants $a$($\AA$), tetragonality $c/a$, unit cell volume $\Omega_0$($\AA^3$), and atomic displacements $dz$ within different functionals for \BTO\ and \PTO. $dz$ relative to Ba(Pb) are given in fractions of c lattice constants. The results of the calculations including the Ti pseudopotential are shown in brackets.}
 \begin{tabular*}{1.0\textwidth}%
    {@{\extracolsep{\fill}}cccccccccccccccc}
\hline\hline
     & \multicolumn{7}{c}{BaTiO$_3$} & & \multicolumn{7}{c}{PbTiO$_3$} \\  
     
                & LDA & \multicolumn{2}{c}{GGA} & \multicolumn{2}{c}{HYBRYDS } & HF & EXP. & & LDA & \multicolumn{2}{c}{GGA} & \multicolumn{2}{c}{HYBRYDS } & HF & EXP \\  
                &  & PBE & WC & B3LYP & B1 &  &  & &  & PBE & WC & B3LYP & B1 &  &  \\  
\hline
\multicolumn{3}{l}{Cubic phase}     &     &     &     & &     &     &     &     &     &     &     &     &     \\
      $a$ &3.958&4.035&3.990&4.036&4.008&4.016& 4.00 & &3.894&3.958&3.923&3.958&3.932&3.933&3.93$^f$ \\ 
                &(3.958)&(4.036)&(3.990)&(4.036)&(4.008)&(4.011)& &    &(3.896)&(3.956)&(3.921)&(3.955)&(3.929)&(3.927)&    \\
                & 3.96$^a$ & 4.0$3^a$ &     & 4.04$^a$ &     & 4.01$^a$ & &    &  3.93$^a$ & 3.96$^a$ & 3.933$^b$ & 3.96$^a$ &    &  3.94$^a$   &   \\
                & 3.951$^c$ & 4.023$^c$ &     &     &     &     &     & & 3.894$^c$ & 3.971$^c$ &     &     &     &     &   \\
                & 3.937$^d$ & 4.027$^d$ &     &     &     &     &     & & 3.890$^e$ &     &     &     &     &     &   \\
                & 3.946$^e$ &      &     &     &     &     &     &     &     &  &     &     &     &     &   \\

\hline
\multicolumn{3}{l}{Tetragonal phase}     &     &     &     &     &     &  &     &     &     &     &     &     &     \\
         $a$ & 3.954 & 4.013 & 3.982 & 3.996 & 3.987 & 3.996 & 3.986$^g$ & & 3.872 & 3.834 & 3.870 & 3.819 & 3.812 & 3.798 & 3.88$^f$  \\ 
                  &(3.951)&(4.009)&(3.978)&(3.988)&(3.982)&(3.977)&       &    &(3.862)&(3.828)&(3.840)&(3.817)&(3.805)&(3.786)&           \\
$c/a$        & 1.006 & 1.035 & 1.012 & 1.066 & 1.036 & 1.037 & 1.010$^g$ & & 1.041 & 1.221 & 1.086 & 1.277 & 1.224 & 1.277 & 1.071$^f$ \\ 
                 &(1.011)&(1.043)&(1.018)&(1.080)&(1.045)&(1.064)&      &     &(1.054)&(1.236)&(1.131)&(1.287)&(1.238)&(1.293)&           \\
$\Omega_0$ & 62.2 & 66.9 & 63.9 & 68.0 & 65.7 & 66.2 & 64.0$^g$ & & 60.4 & 68.8 & 62.9 & 71.1 & 67.8 & 70.0 & 62.6$^f$ \\
                &(62.3)&(67.2)&(64.1)&(68.5)&(66.0)&(67.0)&      &    &(60.7)&(69.3)&(64.1)&(71.5)&(68.2)&(70.2)&          \\
 $dz_{\rm Ti}$       & 0.011 & 0.018 & 0.013 & 0.019 & 0.019 & 0.018 & 0.015$^g$ & & 0.037 & 0.062 & 0.044 & 0.076 & 0.063 & 0.076 & 0.040$^h$  \\
                 &(0.013)&(0.019)&(0.016)&(0.019)&(0.020)&(0.021)&       &    &(0.037)&(0.064)&(0.047)&(0.079)&(0.065)&(0.079)&              \\
$dz_{\rm O_{\parallel}}$        & -0.014 & -0.039 & -0.022 & -0.057 & -0.040 & -0.040 & -0.023$^g$ & & 0.090 & 0.189 & 0.121 & 0.223 & 0.192 & 0.225 & 0.112$^h$ \\
                 &(-0.018)&(-0.043)&(-0.025)&(-0.065)&(-0.045)&(-0.056)&        &    &(0.098)&(0.195)&(0.141)&(0.229)&(0.199)&(0.233)&           \\
$dz_{\rm O_{\perp}}$        & -0.009 & -0.022 & -0.013 & -0.031 & -0.022 & -0.018 & -0.014$^g$ & & 0.106 & 0.178 & 0.133 & 0.198 & 0.181 & 0.199 & 0.112$^h$  \\
                 &(-0.012)&(-0.025)&(-0.016)&(-0.036)&(-0.024)&(-0.027)&       &     &(0.114)&(0.184)&(0.148)&(0.203)&(0.186)&(0.204)&            \\
\hline\hline\\[-20pt]
\end{tabular*}
\end{center}
Optimized cubic lattice constants from: $^a$Ref.~\cite{Piskunov2004}, $^b$Ref.~\cite{Wu2006}, $^c$Ref.~\cite{Wu2004}, $^d$Ref.~\cite{Tinte1998},$^e$Ref.~\cite{King1994}.$^f$Cubic and tetragonal lattice constants of \PTO\ extrapolated to 0 K from Ref.~\cite{Mabud1979}. $^g$Room temperature data from Ref.~\cite{Shirane1957}. $^h$Room temperature data from Ref.~\cite{Shirane1956}
\end{minipage}
\end{table}    
\end{widetext}

First, we calculated the Born effective charges $Z^*$ of the optimized cubic phases. The Born effective charges were obtained computing the electronic polarization for small individual atomic displacements along the $z$-axis with O$_{\parallel}$ and O$_{\perp}$ atoms located in the $z$=0 and $z$=0.5 planes, respectively (see Fig.~\ref{Strucfig}a). $Z^*$ values obtained using different functionals are shown in Table~\ref{FEproptable}. The LDA, GGA-PBE, GGA-WC, B3LYP, and B1 render very similar results for the effective charges, whereas $Z^*_{\rm Ti}$ and $Z^*_{\rm O_{\parallel}}$ obtained using the HF scheme are much smaller than those given by the other functionals. At the LDA level the $Z^*$ values are comparable with those of the previous pseudopotential calculations.\cite{Ghosez1998a} For the different functionals which we have considered, $Z^*_{\rm Ti}$ and $Z^*_{\rm O_{\parallel}}$ have a roughly linear dependence with the band gaps $E_g$, decreasing with $E_g$ values. In comparison to these results, we note that the pseudo-SIC method reported in Ref.~\cite{Pilippetti2003} which provides a gap comparable to the experiment yields surprisingly small values close to HF. For the calculations including Ti pseudopotentials, $Z^*_{\rm Ba/Pb}$ and $Z^*_{\rm O_{\perp}}$ do not change significantly. There is only a small increase in the values of $Z^*_{\rm Ti}$ and $Z^*_{\rm O_{\parallel}}$, suggesting that the HF Ti pseudopotential will lead to an overestimation of the polarization of the FE phases.

Then, the spontaneous polarization $P_s$ was also calculated using the Berry phase approach~\cite{King1993, Resta1994} for the fully optimized tetragonal phases corresponding to each functional (see Table~\ref{FEproptable}). Because GGA-PBE, B3LYP, B1 and HF  yield an erroneous super-tetragonal structure they overestimate the experimental values of $P_s$, which are 27~C/m$^2$ for \BTO~\cite{Wieder1955} and 0.5-1.0~C/m$^2$ for \PTO~\cite{Lines1977}. The LDA is found to underestimate the experimental value of $P_s$ in the case of \BTO\, and to produce a more accurate result for \PTO. The GGA-WC, on the other hand, renders $P_s$ values that are in acceptable agreement with experiment for both \BTO\ and \PTO. The $P_s$ values obtained with Ti pseudopotentials are typically larger than those obtained including all Ti electrons, which is consistent with the overestimations of the tetragonality and Born effective charges. The largest increase for \BTO\ corresponds to the HF results ($\sim$120~mC/m$^2$), whereas for \PTO\ the largest increase is given by the GGA-WC ($\sim$130~mC/m$^2$).      

It is interesting to consider the energy difference $\Delta E_{t}$ between the paraelectric cubic and FE tetragonal phases, as such a quantity strongly correlates with (and can be expected to be roughly proportional to) the FE transition temperature. The cubic-tetragonal phase transition occurs at 393~K (766~K) for \BTO (\PTO), which corresponds to an energy of 34~meV (66~meV). Hence, for \BTO\ the GGA-PBE, B3LYP, and B1 render $\Delta E_{t}$'s that are larger than our reference value, while the LDA, GGA-WC and HF schemes give smaller values. For \PTO, the $\Delta E_{t}$ values obtained using the GGA-PBE, B3LYP, B1, and HF schemes are significantly larger than what one would expect. These results seem to be consistent with the LDA-based first-principles effective-hamiltonian studies in the literature,~\cite{Zhong94, Wagmare1997} which predict transition temperatures that are usually lower than the experimental ones (by as much as 100~K in the case of \BTO\ and \PTO). Our results indicate that the situation might be improved by using functionals that perform better than the LDA in what regards the description of the energetics of the system.

\begin{widetext}
\begin{table}[t]
\begin{minipage}[b]{1.0\textwidth}
\begin{center}
\caption{\label{FEproptable} Born effective charges $Z^*$ calculated at theoretical cubic lattice constants , macroscopic polarization $P$ of the fully relaxed tetragonal phases, and the total energy difference $\Delta E_{t}$ between the optimized cubic and tetragonal phases of \BTO\ and \PTO.  The results including Ti pseudopotentials are shown in brackets.}
 \begin{tabular*}{1.0\textwidth}%
    {@{\extracolsep{\fill}}ccccccccccccccc}
\hline\hline
     & \multicolumn{7}{c}{BaTiO$_3$} & \multicolumn{7}{c}{PbTiO$_3$} \\  
     
                 & LDA & \multicolumn{2}{c}{GGA} & \multicolumn{2}{c}{HYBRYDS } & HF & Exp. &  LDA & \multicolumn{2}{c}{GGA} & \multicolumn{2}{c}{HYBRYDS } & HF & Exp. \\  
                 &  & PBE & WC & B3LYP & B1 &  &  &  & PBE & WC & B3LYP & B1 &   &  \\  
\hline
 \multicolumn{3}{l}{Cubic phase}     &     &     &     &     &     &     &     &     &     &     &     &     \\
     $Z^*_{\rm Ba(Pb)}(|e|)$ & 2.76 & 2.73 & 2.76 & 2.68 & 2.72 & 2.64 &        & 3.84 & 3.87 & 3.85 & 3.78 & 3.83 & 3.67 &   \\ 
                         &(2.74)&(2.71)&(2.73)&(2.67)&(2.70)&(2.64)&        &(3.80)&(3.85)&(3.83)&(3.77)&(3.81)&(3.31)&  \\
     $Z^*_{\rm Ti} (|e|)$    & 7.11 & 7.17 & 7.10 & 6.93 & 6.94 & 5.74 &        & 6.96 & 7.02 & 6.70 & 6.81 & 6.79 & 5.82 &   \\
                         &(7.29)&(7.38)&(7.31)&(7.15)&(7.17)&(5.98)&        &(7.15)&(7.32)&(7.17)&(7.01)&(7.02)&(6.06)&  \\
     $Z^*_{\rm O_{\parallel}} (|e|)$    & -5.65 & -5.73 & -5.66 & -5.35 & -5.55 & -4.41 &  & -5.77 & -5.83 & -5.81 & -5.58 & -5.60 & -4.43 &  \\
                         &(-5.79)&(-5.91)&(-5.82)&(-5.72)&(-5.75)&(-4.63)&  &(-5.91)&(-6.02)&(-5.95)&(-5.77)&(-5.81)&(-4.64)&  \\
      $Z^*_{\rm O_{\perp}} (|e|)$   & -2.12 & -2.08 & -2.10 & -2.04 & -2.05 & -1.99 &  & -2.51 & -2.53 & -2.52 & -2.50 & -2.51 & -2.53 &   \\ 
                         &(-2.12)&(-2.10)&(-2.11)&(-2.05)&(-2.06)&(-1.99)&  &(-2.52)&(-2.53)&(-2.53)&(-2.50)&(-2.51)&(-2.47)&  \\

  \multicolumn{3}{l}{Tetragonal phase}     &     &     &     &     &     &     &     &     &     &     &     &     \\
        $P$(C/m$^2$)      & 0.20 & 0.39 & 0.26 & 0.48 & 0.39 & 0.33 & 0.27$^a$ & 0.78 & 1.29 & 0.98 & 1.40 & 1.31 & 1.39 & 0.5-1$^b$  \\ 
                         &(0.24)&(0.43)&(0.31)&(0.53)&(0.44)&(0.45)&   &(0.85)&(1.32)&(1.11)&(1.43)&(1.35)&(1.43)&   \\
       $\Delta E_t (meV)$    & 11.6 & 40.4 & 14.3 & 82.1 & 48.2 & 20.9 &  & 32.6 & 248.4 & 99.7 & 473.7 & 284.6 & 454.4 &    \\ 
                         &(4.7 )&(49.3)&(14.5)&(99.6)&(59.0)&(86.2)&  &(67.7)&(292.4)&(115.5)&(525.6)&(331.4)&(587.7)&  \\
\hline\hline
\end{tabular*}
\end{center}
Experimental data from: $^a$Ref.~\cite{Wieder1955} and $^b$Ref.~\cite{Lines1977}.
\end{minipage}
\end{table}    
\end{widetext}

\subsection{Summary}

The results discussed so far clearly suggest that only the hybrid functionals can provide an accurate description of the electronic structure of FE materials, as the LDA and GGA typically underestimate the band gap by about a factor of two and HF calculations usually render large overestimations. At the same time, and unfortunately, these hybrid functionals yield erroneous predictions for the atomic structure of the FE phases, as best exemplified by the super-tetragonality problem that also pertains to the GGA-PBE scheme. In fact, we find that only the GGA-WC, and the LDA to a lesser extent, provide an accurate description of the structural properties of \BTO\ and \PTO. As a result, we find that none of the available functionals is able to describe with acceptable accuracy both the electronic and structural properties of these systems.

\begin{table}[h]
\begin{center}
\caption{\label{B3LYPTable} Fully relaxed lattice constants, tetragonality $c/a$, unit cell volume $\Omega_0$, and indirect band gap E$_g^i$ [$R-\Gamma$] of tetragonal \BTO\ for different values of the Becke's mixing parameters. The LDA, B1 and  B1-WC (with the exact exchange A=0.16 and A=0.25), and experimental values are also included.}
 \begin{tabular*}{0.48\textwidth}%
    {@{\extracolsep{\fill}}cccccccc}
\hline\hline
         & $A$ & $B$ & $C$  & $a (\AA)$ & $c/a$ & $\Omega_0 (\AA^3)$ & $E_g^i (eV)$ \\
\hline
LDA   &0  & 0  & 0   & 3.954& 1.006& 62.2& 2.10  \\
      &0.2&   0&    0& 3.948& 1.015& 62.4& 3.77  \\
      &0.2&0.9 &    0& 3.976& 1.138& 71.5& 3.85  \\
      &0.2&   0& 0.81& 3.934& 1.005& 61.2& 3.73  \\
B3LYP &0.2& 0.9& 0.81& 3.996& 1.066& 68.0& 3.80  \\
B1    &0.16& 1 & 1   & 3.987& 1.036& 65.7& 3.45  \\
B1    &0.25& 1 & 1   & 3.977& 1.032& 64.9& 4.26  \\
B1-WC &0.16& 1 & 1   & 3.962& 1.015& 63.2& 3.44  \\
B1-WC &0.25& 1 & 1   & 3.954& 1.015& 62.7& 4.24  \\
Exp.\cite{Shirane1957} &-&-&-& 3.986& 1.01& 64.0& 3.4~\cite{Wemple1970} \\
\hline\hline
\end{tabular*}
\end{center}
\end{table}%

\section{B1-WC hybrid functional}

\subsection{Origin of B3LYP and B1 failure}

In order to understand the origin of super-tetragonality given by the B3LYP functional, we have performed for \BTO\ calculations using different values of the three Becke's mixing parameters (see Eq.~\ref{Eq1}). The results are summarized in Table~\ref{B3LYPTable}.

First we note that the percentage of exact exchange ($A$ parameter) strongly affects the results of the band gaps and $c/a$ ratios, being responsible for the overestimate of the tetragonality obtained within HF. Yet, if we fix $A$ at the value yielding accurate band gaps ($A\approx 0.2$), it becomes clear that the main cause for the B3LYP super-tetragonality is Becke's GGA exchange contribution ($B$ parameter), which also leads to a relatively large value for the unit cell volume. Indeed, for $B$ parameter values in the interval between 0.0 and 0.9, and no correlation contribution ($C$=0) the tetragonality (resp. unit cell volume) increases by $\sim12$\% (resp. $\sim15$\%). It is also worth to note that the value of $B$ does not affect the calculated band gap significantly. Finally, the GGA correlation ($C$ parameter) has a small influence on the band gap, tetragonality and unit cell volume. For $C$ parameter values between 0.0 and 0.81, and no exchange contribution ($B$=0), the tetragonality (resp. unit cell volume) decreases only by $\sim 1$\% (resp. $\sim 2$\%). 

This analysis clearly demonstrates that the failures of the B3LYP and B1 functionals, that are, the super-tetragonality and the severe overestimation of the unit cell volume, are essentially caused by Becke's GGA exchange part, and that the tuning of the GGA correlation part cannot compensate for such inaccuracies.

\subsection{B1-WC hybrid functional}

The above discussion suggests that the super-tetragonality problem associated to B3LYP and B1 might be overcome by improving for the GGA exchange part. Since, as highlighted above, GGA-WC describes the structural properties of \BTO\ and \PTO\ quite accurately, we have built a so called B1-WC hybrid functional by mixing exact exchange with  GGA-WC following the recent B1 Becke's mixing scheme. This scheme only includes the mixing parameter $A$ (see Eq.~\ref{Eq1}), with values ranging from 0.16 to 0.28 depending of the choice of the GGA.\cite{Becke1996,Ernzerhof1996}  Since A monitor the fraction of exact exchange, the amplitude of the gap is directly linked to its value and constitutes good indicator for fixing its value. We have performed calculations with $A$=0.16 and $A$=0.25 for tetragonal phases of \BTO\ (see Table~\ref{B3LYPTable}). From Table~\ref{B3LYPTable}, it is clear that going above $A$=0.16 percentage produces unrealistically large band gaps for \BTO. Moreover, B1-WC with $A$=0.16 also gives the best structural properties. For comparison the calculations within B1 functional are also included in Table~\ref{B3LYPTable}.


In order to demonstrate the universality of this choice for the class of ABO$_3$ ferroelectrics, we have performed full relaxation calculations of the cubic phases for different compounds. These results are shown in Table~\ref{B1WCCubicTable}. In all the compounds considered E$_g^i$(E$_g^d$) occurs along $R-\Gamma$($\Gamma-\Gamma$) direction of the cubic Brillouin zone except for \PTO\ and \PZO. For \PZO, E$_g^i$(E$_g^d$) is along $R-X$($X-X$), whereas for \PTO, E$_g^i$(E$_g^d$) is along $X-\Gamma$($\Gamma-\Gamma$). The agreement with experiment for the band gaps and lattice constant is very good for all the considered compounds except \PTO\ and \PZO. For \PTO\ and \PZO, we have also performed calculations using Durand pseudopotential for Pb.\cite{Nizam1988} Changing the Pb pseudopotential does not significantly affect the values of the band gaps. In the case of \PTO, E$_g^i$ increases by 0.2 eV, whereas for \PZO, E$_g^i$ decreases by 0.06 eV. Trying to understand the origin of this discrepancy, it is worth to notice that the band gaps of \PTO\ and \PZO\ were not directly measured but extrapolated to 0 K. The reported values strongly depend of the method and the choice of parameters used to extrapolate the high temperature data.\cite{Zametin1984}

Since B1-WC with the exact exchange parameter $A$=0.16 gives good structural and electronic properties for the cubic phases of the considered compounds, further, we have calculated the other structural and electronic properties of the cubic and tetragonal phases of \BTO\ and \PTO\ (see Table~\ref{B1WCTable}). The B1-WC electronic band structure of \BTO\, obtained from the Ti all-electron calculation, is shown in Fig.~\ref{Bndfig}c. The result is similar to that of the $GW$ calculation (Fig.~\ref{Bndfig}d), except the position of the Ba~5$p$ states which are higher in energy by $\sim$1.5 eV. We notice that even the degeneracy of the CB at the $X$ point, which erroneously appeared in the B3LYP calculation, is properly lifted up. This degeneracy is lifted up at the level of B1 functional, which gives very similar band structure results with those of B1-WC. For the Ti all-electron calculations, we obtain structural properties in very good agreement with experiment for the cubic and tetragonal phases of both \BTO\ and \PTO. As compared with the GGA-WC results, B1-WC renders cubic lattice constants that are smaller by 0.5$\%$ for both \BTO\ and \PTO. For the tetragonal phase the unit cell volume is smaller by $\sim1.1\%$ (resp. $\sim0.8\%$), while the tetragonality increases by $\sim0.3\%$ (resp. $\sim1\%$) for \BTO(resp. \PTO).

\begin{table}[t]
\begin{center}
\caption{\label{B1WCCubicTable}Fully optimized lattice constants, indirect E$_g^i$ and direct E$_g^d$ band gaps of different cubic phases within B1-WC hybrid functional. The experimental values are also shown for comparison. The results with Ti pseudopotentials are shown in brackets for \BTO\ and \PTO.}
 \begin{tabular*}{0.48\textwidth}%
    {@{\extracolsep{\fill}}ccccccc}
\hline\hline
                & \multicolumn{2}{c}{$a$(\AA)} & \multicolumn{2}{c}{$E_g^i (eV)$} & \multicolumn{2}{c}{$E_g^d (eV)$} \\ 
                 & B1-WC & Exp. & B1-WC & Exp. & B1-WC & Exp.  \\

\hline
BaTiO$_3$        & 3.971 & 4.00 & 3.39 & 3.20\cite{Wemple1970} & 3.45 &   \\  
                 &(3.97) &      &(3.16)&                       &(3.21)&   \\
PbTiO$_3$        & 3.901 & 3.93\cite{Mabud1979} & 2.68 & 3.39\cite{Zametin1984} & 4.23 &   \\
                 &(3.90) &                      &(2.43)& 3.40\cite{Peng1992} &(3.98)&   \\
SrTiO$_3$        & 3.880 & 3.890$^a$\cite{Hellwege1969}& 3.57 & 3.25\cite{Benthem2001} & 3.91 & 3.75\cite{Benthem2001} \\
CaTiO$_3$        & 3.834 & 3.836$^b$\cite{Yoshiasa2001} & 3.59 & 3.57\cite{Ueda1998}  & 4.04 &   \\ 
BaZrO$_3$        & 4.195 & 4.192\cite{Yamanaka2003} & 5.24 & 5.33\cite{Robertson2000} & 5.41 &   \\   
PbZrO$_3$        & 4.148 & 4.130$^c$\cite{Sawaguchi1953} & 3.65 &    & 3.40 & 3.83\cite{Zametin1984} \\  
SrZrO$_3$        & 4.138 & 4.109\cite{Smith1960} & 5.51 & 5.60\cite{Lee2003}          & 5.77 &   \\
CaZrO$_3$        & 4.111 &                       & 5.35 &                             & 5.65 &   \\
KTaO$_3$         & 3.971 & 3.988$^d$\cite{Adachi1972} & 3.41 & 3.57\cite{Frova1967} & 4.07 & 4.40\cite{Frova1967}  \\
                 &       &      &      & 3.64\cite{Jellison2006}&    & 4.35\cite{Jellison2006}\\  

\hline\hline
\end{tabular*}
\end{center}
$^a$Extrapolated to 0 K. $^b$At 600 K. $^c$Extrapolated to 0 K. $^d$Room temperature value. $^e$Extrapolated to 0 K.   
\end{table}

\begin{table}[h]
\begin{center}
\caption{\label{B1WCTable}Born effective charges $Z^*$ of cubic phases, fully optimized tetragonal lattice constants, indirect E$_g^i$ and direct E$_g^d$ band gaps of tetragonal phases, atomic displacements $dz$ in fractions of c lattice constant, macroscopic polarization $P$, and energy difference $\Delta E_t$ between cubic and tetragonal phases for \BTO\ and \PTO\ within B1-WC hybrid functional. The results with Ti pseudopotentials are shown in brackets. The experimental values are also given.  }
 \begin{tabular*}{0.48\textwidth}%
    {@{\extracolsep{\fill}}ccccc}
\hline\hline
                & \multicolumn{2}{c}{BaTiO$_3$} &  \multicolumn{2}{c}{PbTiO$_3$} \\
                & B1-WC & Exp. &  B1-WC & Exp. \\ 
\hline
 \multicolumn{3}{l}{Cubic phase}  &  & \\
$Z^*_{\rm Ba(Pb)} (|e|)$ & 2.74 &    &  3.83   &    \\
                    &(2.72)&   &(3.89)&  \\
$Z^*_{\rm Ti} (|e|)$ & 7.08 &    &  6.81   &   \\
                &(7.11)&    &(6.89)   &   \\
$Z^*_{\rm O_{\parallel}} (|e|)$ & -5.57 &   & -5.62   &  \\
                &(-5.68)&   &(-5.76)  &   \\
$Z^*_{\rm O_{\perp}} (|e|)$ & -2.12 &   &  -2.51  &   \\
                &(-2.08)&   & -2.51   &  \\
  \multicolumn{3}{l}{Tetragonal phase}  &  & \\
$a$(\AA) & 3.962 & 3.986~\cite{Shirane1957}  & 3.846  & 3.88~\cite{Mabud1979} \\
          &(3.957)&   &(3.810)&   \\
$c/a    $ & 1.015 & 1.010~\cite{Shirane1957}  & 1.097  & 1.071~\cite{Mabud1979} \\
          &(1.022)&   &(1.154)&  \\
$\Omega_0 ($\AA$^3)$  & 63.2 & 64.0~\cite{Shirane1957}  & 62.4  & 62.6~\cite{Mabud1979} \\
                  &(63.3)&   &(63.9)&  \\
$E_g^i (eV)$& 3.44  & 3.40~\cite{Wemple1970}     & 2.83   & 3.60\cite{Zametin1984}  \\
           &(3.22)&   &(2.66)&  \\
$E_g^d (eV)$& 3.73  &       &  4.94   &  \\
           & (3.57) &       & (4.76) &   \\
$dz_{\rm Ti}$ & 0.015 & 0.015~\cite{Shirane1957}  & 0.046 & 0.040~\cite{Shirane1956} \\
          & (0.017) &    &(0.050) &  \\
$dz_{\rm O_{\parallel}}$ & -0.024 & -0.023~\cite{Shirane1957}  & 0.129 & 0.112~\cite{Shirane1956} \\
          &(-0.029) &   &(0.154) &  \\
$dz_{\rm O_{\perp}}$ & -0.014 & -0.014~\cite{Shirane1957} & 0.140 & 0.112~\cite{Shirane1956} \\
          &(-0.017) &   &(0.158)  &  \\
$P$(C/m$^2$) & 0.28 & 0.27~\cite{Wieder1955}  &  1.03  &0.5-1.0~\cite{Lines1977}  \\ 
            & (0.33) &  & (1.19) &    \\ 
$\Delta E_t (meV)$& 24.0 &  &  110.6  &  \\ 
            & (24.7) &  & (162.0) &    \\ 
\hline\hline
\end{tabular*}
\end{center}
\end{table}

Concerning the FE properties, we obtain very good $P_s$ values for both \BTO\ and \PTO, the B1-WC Born effective charges being comparable to those obtained with the other functionals (except HF). The energy differences $\Delta E_t$ between the cubic and tetragonal phases are significantly larger than those obtained from LDA calculations. According to the discussion above, in the case of \BTO\ such an increase probably implies that a B1-WC--based thermodynamic calculation will render transition temperatures in significantly better agreement with experiment than those obtained from the LDA. In the case of \PTO\, however, the increase in $\Delta E_t$ seems rather large, and will probably overcorrect the transition temperatures.

As a final check, we also investigated to which extent the new B1-WC hybrid functional performs in predicting the vibrational spectra. We have calculated the phonon frequencies at the zone center ($\Gamma$ point) for cubic phases and compared these values with the corresponding values within LDA and GGA-WC. Phonon frequency calculations in the literature are performed at the optimized or experimental lattice constants; thus, for a better comparison, here we report both types of calculations. 

In the cubic perovskite structure at the $\Gamma$ point there are 12 optical modes: three triply-degenerated modes of $F_{1u}$ symmetry and one triply-degenerated silent mode of $F_{2u}$ symmetry. Transverse frequencies of these different modes are reported in Table~\ref{FreqTable}, together with experimental values in the case of 
\BTO.\cite{Luspin1980}
 
\begin{table}[h]
\begin{center}
\caption{\label{FreqTable}Phonon frequencies ($cm^{-1}$) at the $\Gamma$ point for cubic phases of \BTO\ and \PTO\ calculated at optimized and experimental lattice constants. The results with Ti pseudopotentials are shown in brackets. The experimental values for \BTO\ are also included.   }
  \begin{tabular*}{0.48\textwidth}%
     {@{\extracolsep{\fill}}cccccccc}
\hline\hline
                 & \multicolumn{4}{c}{BaTiO$_3$} &  \multicolumn{3}{c}{PbTiO$_3$} \\
                 & LDA & WC & B1-WC & Exp. & LDA & WC & B1-WC \\ 
\hline
\multicolumn{3}{l}{$a=a_{opt}$}  &  &  &  &  &  \\
$F_{1u}(TO1)$            & 75i & 128i & 145i & soft & 127i & 132i & 146i \\
                 & (157i) & (192i) & (213i) &  & (148i) & (163i) & (196i) \\
$F_{1u}(TO2)$            & 193 & 186 & 195 & 182$^a$ & 145 & 141 & 138 \\
                 &(193)  & (186) & (195) &  & (128) & (115) & (120) \\
$F_{1u}(TO3)$            & 480 & 469 & 482 & 482$^a$ & 515 & 510 & 513 \\
                 & (474) & (463) & (476) &  & (509) & (495) & (506) \\
$F_{2u}$         & 286 & 282 & 299 & 306$^b$ & 219 & 211 & 231 \\
                 & (283) & (280) & (298) &  & (214) & (208) & (229)  \\
\multicolumn{3}{l}{$a=a_{exp}$}  &  &  &  &  &  \\
$F_{1u}(TO1)$            & 187i & 155i & 208i & soft & 139i & 135i & 155i \\
                 & (218i) & (204i) & (255i) &  & (170i) & (168i) & (192i) \\
$F_{1u}(TO2)$            & 181 & 183 & 186 & 182$^a$ & 132 & 137 & 127 \\
                 & (180) & (183) & (185) &  & (107) & (109) & (96) \\
$F_{1u}(TO3)$            & 456 & 462 & 468 & 482$^a$  & 491 & 503 & 497  \\
                 & (453) & (458) & (463) &  & (479) & (487) & (481) \\
$F_{2u}$         & 281 & 281 & 295 & 306$^b$ & 216 & 210 & 229  \\
                 & (278) & (279) & (294) &  & (213) & (208) &  (228) \\
\hline\hline
\end{tabular*}
\end{center}
\begin{flushleft}
Experimental data from: $^a$Ref.~\cite{Luspin1980} $^b$This value was measured in tetragonal phase.
\end{flushleft}
\end{table}%

At the optimized lattice constant, for \BTO, the best agreement with the experimental data is achieved with the B1-WC hybrid functional. It is for the $F_{1u}(TO1)$ ferroelectric soft mode that the dispersion of the results is the largest, which is consistent with the well known fact that the FE soft-mode frequency is strongly dependent on the unit cell volume. Consequently, the dispersion in the computed soft-mode frequencies reduces significantly when the calculations are made at the experimental lattice constant. For \PTO, although we cannot compare with experimental data, we observe that B1-WC provides results comparable to the other functionals, all of them being in rather close agreement.  We finally notice that the soft-mode (and to a lower extent the $F_{1u}(TO2)$ mode of \PTO) is strongly affected by the use or not of the Ti HF pseudopotential. This further highlights the very delicate nature of the ferroelectric instability and the necessity to be very careful in order to reproduce it properly.

\section{Conclusions}

In summary, our electronic structure calculations of prototypical FE crystals \BTO\ and \PTO\, using the most popular exchange-correlation functionals, show that it is difficult to obtain simultaneously good accuracy for structural and electronic properties. On the one hand, all the usual DFT functionals (LDA, GGA-PBE, GGA-WC) reproduce the structural properties with various degrees of success, the recently introduced GGA-WC being by far the most accurate, but severely underestimate the band gaps. On the other hand, the B3LYP and B1 hybrid functionals correct the band-gap problem, but overestimate the volumes and atomic distortions giving a super-tetragonality comparable to that rendered by the GGA-PBE for the tetragonal phases.

We found that the super-tetragonality inherent to  B3LYP and B1 calculations is mostly associated to the GGA exchange part and, to a lesser extent, to the exact-exchange contribution. To bypass this problem, we have proposed a different B1 hybrid functional obtained by combining the GGA-WC functional with a small percentage of exact exchange ($A$=0.16). With this B1-WC, we have obtained very good structural, electronic and ferroelectric properties as compared to experimental data for \BTO\ and \PTO. Very good agreement with experiment is obtained for the lattice constants and the band gaps for cubic phases of other perovskite compounds, except the band gaps of \PTO\ and \PZO. This different B1-WC functional thus opens the door to a better description of the optical properties of ferroelectrics and of metal/FE interfaces for which the tetragonal distortion of the crystalline cell, the atomic displacements, the electronic structure, and in particular the band gap, have to be accurately described simultaneously.

\begin{acknowledgments}

PhG acknowledges financial support from the VolkswagenStiftung (I/77 737), the Interuniversity Attraction Poles Program - Belgian State - Belgian Science policy (P6/42), the  FAME European Network of Excellence and the MaCoMuFi European Strep project. J\'I\ acknowledges financial support from the Spanish Ministry of Science and Education (FIS2006-12117-C04-01 and CSD2007-00041) and the Catalan Government (SGR2005-683). We made use of the facilities of the Barcelona Supercomputing Center(BSC-CNS).

\end{acknowledgments}


\begin{thebibliography}{99}

\bibitem{Waser}
R. Waser, {\it Nanoelectronics and information technology : advanced electronic materials and novel devices} (Wiley-VCH, Weinheim, 2003). 

\bibitem{Scott}
J. F. Scott,  {\it Ferroelectric memories} (Springer Verlag, Berlin, 2000).

\bibitem{Rabe-Ghosez}
K. M. Rabe and Ph. Ghosez, in {\it Modern Ferroelectrics}, Ed. by K. M. Rabe, C. H. Ahn and J.-M. Triscone,  Topics Applied Physics {\bf 105}, 111-166 (Springer-Verlag, Berlin, 2007).

\bibitem{Ghosez06}
Ph. Ghosez et J. Junquera, in {\it Handbook of theoretical and computational nanotechnology}, Ed. by M. Rieth and W. Schommers, vol. {\bf 9},  p. 623-728 (ASP, Stevenson Ranch CA, 2006).

\bibitem{Zimmer02}
M. Zimmer, J. Junquera, and P. Ghosez, in {\it Fundamental Physics of Ferroelectrics 2002 }, Ed. by R. E. Cohen, vol. {\bf 626}, p. 232-241 (AIP Conference Proceedings, Melville, New York, 2002).


\bibitem{Junquera03}
J. Junquera, M. Zimmer, P. Ordejon, and P. Ghosez, \Journal{Phys. Rev. B}{67}{155327}{2003}.  

\bibitem{PBE} 
J. P. Perdew, K. Burke, and M. Ernzerhof, \Journal{ Phys. Rev. Lett.}{77}{3865}{1996}.

\bibitem{Wu2004} 
Z. Wu, R. E. Cohen, and J. D. Singh, \Journal{ Phys. Rev. B}{70}{104112}{2004}.

\bibitem{Wu2006} 
Z. Wu and R. E. Cohen, \Journal{ Phys. Rev. B}{73}{235116}{2006}.
 
\bibitem{Gunnarsson79} 
O. Gunnarsson, M. Jonson, and B. I. Lundqvist, \Journal{Phys. Rev. B}{20}{3136}{1979}.

\bibitem{Singh97} 

D. J. Singh, \Journal{Ferroelectrics}{194}{299}{1997}.

\bibitem{Martin04}
R. Martin, {\it Electronic Structure Basic Theory and Practical Methods}, p. 44 ( Cambridge University Press, Cambridge, 2004).  

\bibitem{Gonze1995} 
X. Gonze, Ph. Ghosez and R. W. Godby, \Journal { Phys. Rev. Lett.}{74}{4035}{1995}.

\bibitem{Ghosez1997} 
Ph. Ghosez, X. Gonze, and  R. W. Godby, \Journal{ Phys. Rev. B}{56}{12811}{1997}.

\bibitem{Hill1998} 
N. A. Hill and K. M. Rabe, \Journal{ Phys. Rev. B}{59}{8759}{1998}.

\bibitem{Hill2002} 
A. Filippetti and N. A. Hill, \Journal{ Phys. Rev. B}{65}{195120}{2002}.

\bibitem{Shishidou2004} 
T. Shishidou, N. Mikamo, Y. Uratani, F. Ishii, and T. Oguchi, \Journal{ J. Phys.: Condens. Matter}{16}{S5677}{2004}.  

\bibitem{Becke1} A. D. Becke, \Journal{ J. Chem. Phys.}{98}{1372}{1993}.

\bibitem{Becke1993} A. D. Becke, \Journal{ J. Chem. Phys.}{98}{5648}{1993}.

\bibitem{CRYSTAL2005}
R. Dovesi, R. Orlando, B. Civalleri, C. Roetti, V. R. Saunders, and C. M. Zicovich-Wilson, \Journal{ Zeitschrift Fur Kristallographie}{220}{571}{2005}.

\bibitem{Becke1996} 
A. D. Becke, \Journal{ J. Chem. Phys.}{104}{1040}{1996}.

\bibitem{Ernzerhof1996} 
M. Ernzerhof, J. P. Perdew, and K. Burke, in { \it Density Functional Theory}, Ed. by R. Nalewajski (Springer, Berlin, 1996).  

\bibitem{Barone1994} 
V. Barone, \Journal{ Chem. Phys. Lett.}{226}{392}{1994}.

\bibitem{Bausch1995} 
C. W. Bauschlicher, \Journal{ Chem. Phys. Lett.}{246}{40}{1995}.

\bibitem{Baker1995} 
J. Baker, J. Andzelm, M. Muir, and P. R. Taylor, \Journal{ Chem. Phys. Lett.}{237}{53}{1995}.

\bibitem{Tozer1996} 
D. J. Tozer, \Journal{ J. Chem. Phys.}{104}{4166}{1996}.

\bibitem{Neumann1996} 
R. Neumann and N. C. Handy, \Journal{ Chem. Phys. Lett.}{252}{19}{1996}.

\bibitem{Dori2006}
N. Dori, M. Menon, L. Kilian, M. Sokolowski, L. Kronik, and E. Umbach, \Journal{ Phys. Rev. B}{73}{195208}{2006}.

\bibitem{Riley2007}
K. E. Riley and K. M.  Merz, \Journal{ J. Phys. Chem. A}{111}{6044}{2007}.

\bibitem{Martin1997}
R. L. Martin and F. Illas,  \Journal{ Phys. Rev. Lett.}{79}{1539}{1997}.

\bibitem{Bredow2000}
T. Bredow and A. R. Gerson,  \Journal{ Phys. Rev. B}{61}{5194}{2000}.

\bibitem{Muscat2001}
J. Muscat, A. Wander and N. M. Harrison, \Journal{ Chem. Phys. Lett.}{342}{397}{2001}.

\bibitem{Feng2004}
X. Feng and N. M. Harrison, \Journal{ Phys. Rev. B}{70}{092402}{2004}.

\bibitem{Cora2004}
F. Cora, M. Alfredsson, G. Mallia, D. S. Middlemiss, W. C. Mackrodt, R. Dovesi, and R. Orlando, \Journal{ Structure and Bonding }{113}{171}{2004}.

\bibitem{Paier2006}
J. Paier, M. Marsman, K. Hummer, G. Kresse, I. C. Gerber, and J. G. Angyan, \Journal{ J. Chem. Phys.}{124}{154709}{2006}.


\bibitem{Tran2006}
F. Tran, P. Blaha, K. Schwarz, and P. Novak, \Journal{ Phys. Rev. B}{74}{155108}{2006}.

\bibitem{Paier2007}
J. Paier, M. Marsman, and G. Kresse, \Journal{ J. Chem. Phys.}{127}{024103 }{2007}. 

\bibitem{Gerber2007}
I. C. Gerber, J. G. Angyan, M. Marsman, G. Kresse,  \Journal{ J. Chem. Phys.}{127}{054101}{2007}.


\bibitem{Ruiz2003}
E. Ruiz, M. Llunell, and P. Alemany,  \Journal{J. Sol.  Stat. Chem.}{176}{400}{2003}.

\bibitem{Franchini2005}
C. Franchini, V. Bayer, R. Podloucky, J. Paier, and G. Kresse, \Journal{ Phys. Rev. B}{72}{045132}{2005}.

\bibitem{Crespo2006}
R. Grau-Crespo, F. Corà, A. A. Soko, N. H. de Leeuw, and C. R. A. Catlow, \Journal{ Phys. Rev. B}{73}{035116 }{2006}.

\bibitem{Franchini2007} 
C. Franchini, R. Podloucky, J. Paier, M. Marsman, and G. Kresse, \Journal{ Phys. Rev. B}{75}{195128}{2007}. 

\bibitem{Piskunov2004} 
S. Piskunov, E. Heifets, R. I. Eglitis, and G. Borstel, \Journal{ Comp. Mat. Sci.}{29}{165}{2004}.

\bibitem{Harris1974}
J. Harris and R. 0. Jones, \Journal{ J. Phys. F}{4}{1170}{1974}; 0. Gunnarsson
and B. I. Lundqvist, \Journal{ Phys. Rev. B}{13}{4274}{1976}; D. C. Langreth and
J. P. Perdew, \Journal{ Phys. Rev. B}{15}{2884}{1977}; J. Harris, \Journal{ Phys. Rev. A}{29}{1648}{1984}.

\bibitem{Becke1988a}
A. D. Becke, \Journal{ J. Chem. Phys.}{88}{1053}{1988}.

\bibitem{Levy1996} 
M. Levy, N. H. March, and N. C. Handy, \Journal{ J. Chem. Phys.}{104}{1989}{1996}. 

\bibitem{Becke1988} 
A. D. Becke, \Journal{ Phys. Rev. A}{38}{3098}{1988}.

\bibitem{Perdew1991} 
J. P. Perdew, in { \it Electronic Structure of Solids }, edited by P. Ziesche and H. Eschrig ( Akademie Verlag, Berlin, 1991); J. P. Perdew, J. A. Chevary, S. H. Vosko, K. A. Jackson, M. R. Pederson, D. J. Singh, and C. Fiolhais,  \Journal{ Phys. Rev. B}{46}{6671}{1992}.

\bibitem{Lee1988} 
C. Lee, W. Yang, and R. G. Parr, \Journal{ Phys. Rev. B}{37}{785}{1988}.

\bibitem{Perdew1996} 
J. P. Perdew, M. Ernzerhof, and K. Burke, \Journal{ J. Chem. Phys.}{105}{9982}{1996}.

\bibitem{Dirac1930} 
P. A. M. Dirac, \Journal{ Proc. Camb. Phyl. Soc.}{26}{376}{1930}.

\bibitem{Vosko1980} 
S. H. Vosko, L. Wilk, M. Nusair, \Journal{ Can. J. Phys.}{58}{1200}{1980}. 

\bibitem{Catti1991}
M. Catti, R. Dovesi, A. Pavese, and V. R. Saunders, \Journal{ J. Phys. Cond. Matt.}{3}{4151}{1991}.

\bibitem{DovesiZrbasis}
R. Dovesi, Unpublished, { \it http://www.crystal.unito.it/Basis\_Sets/Ptable.html}.

\bibitem{Dovesi1991}
R. Dovesi, C. Roetti, C. Freyria Fava, M. Prencipe, and V.R. Saunders, \Journal{ Chem. Phys.}{156}{11}{1991}.

\bibitem{Bredow2006}
T. Bredow, M.-W. Lumey, R. Dronskowski, H. Schilling, J. Pickardt, M. Lerch, \Journal{ Z. Anorg. Allg. Chem.}{632}{1157}{2006}.

\bibitem{Hedin} 
L. Hedin, \Journal{Phys. Rev.}{139}{A796}{1965};
L. Hedin and S. Lundqvist, \Journal{Solid State Phys.}{ 23}{1}{1969}.

\bibitem{ABINITGW} 
X. Gonze, G. -M. Rignanese, M. Verstraete, J. -M. Beuken, Y. Pouillon, R. Caracas, F. Jollet, M. Torrent, G. Zerah, M. Mikami, Ph. Ghosez,
M. Veithen, J. -Y. Raty, V. Olevano, F. Bruneval, L. Reining, R. Godby,
G. Onida, D. R. Hamann, and D. C. Allan, \Journal{Z. Kristallogr.}{220}{558}{2005}.

\bibitem{Teter} 
M. Teter, \Journal{Phys. Rev. B}{48}{5031}{1993}.

\bibitem{Hybertsen-Godby} 
M. S. Hybertsen and S. G. Louie, \Journal{Phys. Rev. Lett.}{55}{1418}{1985}; \Journal{Phys. Rev. B}{32}{7005}{1985}; \Journal{Phys. Rev. B}{34}{5390}{1986}; R. W. Godby, M. Schl\"uter, and L. J. Sham, \Journal{Phys. Rev. Lett.}{56}{2415}{1986}; \Journal{Phys. Rev. B}{37}{10159}{1988}.

\bibitem{Godby-Needs} 
R. W. Godby and R. J. Needs,  \Journal{Phys. Rev. Lett.}{62}{1169}{1989}.

\bibitem{footnote}
Calculations concerning the rhombohedral phase of BaTiO$_3$ have also been performed and confirm the conclusions reported in the paper.

\bibitem{note-1}
For the calculations which include Ti pseudopotentials, $E^i_g$ and $E^d_g$ values are decreased by $\sim0.2$eV for all functionals except HF. In the case of HF this decrease is larger ($\sim0.4$eV for \BTO\ and $\sim1$eV for \PTO).

\bibitem{Pilippetti2003} 
A. Filippetti and N. A. Hill, \Journal{ Phys. Rev. B}{68}{045111}{2003}.

\bibitem{Wemple1970} 
S. H. Wemple, \Journal{ Phys. Rev. B}{2}{2679}{1970}.

\bibitem{Peng1992} 
C. H. Peng, J. F. Chang, and  S. Desu, in { \it Ferroelectric Thin Films II}, edited by A. I. Kingon, R. E. Myers, and B. Tuttle, MRS Symposia Proceedings No. 243 (Materials Research Society, Pittsburgh, 1992), p. 12.

\bibitem{Shirane1957} 
G. Shirane, H. Danner, and P. Pepinsky, \Journal{ Phys. Rev.}{105}{856}{1957}.

\bibitem{Shirane1956} 
G. Shirane, P. Pepinsky, and B. C. Frazer,  \Journal{ Acta Cryst.}{9}{131}{1956}.



\bibitem{King1994} 
R. D. King-Smith and D. Vanderbilt, \Journal{ Phys. Rev. B}{49}{5828}{1994}. 

\bibitem{Tinte1998} 
S. Tinte, M. G. Stachiotti, C. O. Rodriguez, D. L. Novikov, and N. E. Christensen, \Journal{ Phys. Rev. B}{58}{11959}{1998}.

\bibitem{Hudson1993}
L. T. Hudson, R. L. Kurtz, S. W. Robey, D. Temple, and R. L. Stockbauer, \Journal{ Phys. Rev. B}{47}{1174}{1993}.

\bibitem{Mabud1979} 
S. A. Mabud and A. M. Glazer,  \Journal{ J. Appl. Crystallogr.}{12}{49}{1979}.

\bibitem{Ghosez1998a}
Ph. Ghosez, J.-P. Michenaud, and X. Gonze, \Journal{ Phys. Rev. B}{58}{6224}{1998}.

\bibitem{King1993} 
R. D. King-Smith and D. Vanderbilt, \Journal{ Phys. Rev. B}{47}{1651}{1993};D. Vanderbilt and R. D. King-Smith, \Journal{ Phys. Rev. B}{48}{4442}{1993}.

\bibitem{Resta1994} 
R Resta, \Journal{ Rev. Mod. Phys.}{66}{899}{1994}.

\bibitem{Wieder1955} 
H. H. Wieder, \Journal{ Phys. Rev. }{99}{1161}{1955}.

\bibitem{Lines1977} 
M. E. Lines and A. M. Glass { \it Principles and Applications of Ferroelectrics and Related Materials} ( Oxford University Press, Oxford, 1977), Chap.8.

\bibitem{Zhong94}
W. Zhong, D. Vanderbilt, and K. M. Rabe, \Journal{ Phys. Rev. Lett.}{73}{1861}{1994};W. Zhong, D. Vanderbilt, and K. M. Rabe, \Journal{ Phys. Rev. B}{52}{6301}{1995}.

\bibitem{Wagmare1997}
U. V. Waghmare and K. M. Rabe, \Journal{ Phys. Rev. B}{55}{6161}{1997}.

\bibitem{Nizam1988}
M. Nizam, Y. Bouteiller, B. Silvi, C. Pisani, M. Causa'  and R. Dovesi,
 \Journal{J. Phys. C: Solid State Phys.}{21}{5351}{1988}.

\bibitem{Hellwege1969}
{ \it Ferroelectrics and Related Substances}, Landolt-Bornstein, New Series Group III, Vol. 3, edited by K. H. Hellwege and A. M. Hellwege (Springer-Verlag, Berlin, 1969).

\bibitem{Benthem2001}
K. van Benthem, C. Elsasser, and R.H. French, \Journal{ J. Appl. Phys.}{90}{6156}{2001}.

\bibitem{Yoshiasa2001}
A. Yoshiasa, K. Nakajima,  K-I. Muraia, and M. Okubea, \Journal{ J. Synchrotron Rad.}{8}{940}{2001}.

\bibitem{Ueda1998}
K. Ueda, H. Yanagi, R. Noshiro, H. Hosono, and H. Kawazoe, \Journal{ J. Phys. Cond. Matt.}{10}{3669}{1998}. 

\bibitem{Yamanaka2003}
S. Yamanaka, H. Fujikane, T. Hamaguchi, H. Muta, T. Oyama, T. Matsuda, S. Kobayashi, and K. Kurosaki, \Journal{ J. Alloys Compd.}{359}{109}{2003}.

\bibitem{Robertson2000}
J. Robertson, \Journal{ J. Vac. Sci. Technol. B}{18}{1785}{2000}.

\bibitem{Sawaguchi1953}
E. Sawaguchi, \Journal{ J. Phys. Soc. Jpn.}{8}{615}{1953}.

\bibitem{Smith1960}
A. J. Smith, and A. J. E. Welch, \Journal{Acta Crystallogr.}{13}{653}{1960}.  

\bibitem{Lee2003}
Y. S. Lee, J. S. Lee, T. W. Noh, D. Y. Byun, K. S. Yoo, K. Yamaura, and E. Takayama-Muromachi, \Journal{ Phys. Rev. B}{67}{113101}{2003}. 

\bibitem{Adachi1972}
M. Adachi and A. Kawabata, \Journal{Jpn. J. Appl. Phys.}{11}{1855}{1972}.

\bibitem{Frova1967}
A. Frova and P. J. Boddy, \Journal{Phys. Rev.}{153}{606}{1967}.

\bibitem{Jellison2006}
G. E. Jellison, Jr., I. Paulauskas, L. A. Boatner, and D. J. Singh, \Journal{Phys. Rev. B}{74}{155130}{2006}. 

\bibitem{Zametin1984}
V. I. Zametin,\Journal{ Phys. Stat. Sol. (b)}{124}{625}{1984}. 

\bibitem{Luspin1980}
Y. Luspin, J. L. Servoin, and F. Gervais, \Journal{J. Phys. C }{13}{3761}{1980}.


\end{thebibliography}
\end{document}